\documentclass[sigconf,nonacm]{acmart}

\setcopyright{none}
\copyrightyear{2023}

\usepackage{graphicx}
\DeclareGraphicsExtensions{.pdf,.eps,.png,.jpg}
\graphicspath{{figures/}}
\usepackage{amsmath}
\usepackage[inline]{enumitem}
\usepackage{balance}

\usepackage{booktabs}
\usepackage{tabularx}
\usepackage{xltabular}
\usepackage{multicol}
\usepackage{multirow}
\usepackage{siunitx}
\DeclareSIUnit[number-unit-product = \,]{\USD}{\textsc{USD}}
\DeclareSIUnit{\dollar}{\$}
\DeclareSIPrefix{\million}{\text{ million }}{6}
\DeclareSIPrefix{\billion}{\text{ billion }}{9}

\usepackage{csquotes}
\usepackage{xspace}
\usepackage{caption}
\usepackage{subcaption}
\hypersetup{%
  pdftitle={How Does Connecting Online Activities to Advertising Inferences Impact Privacy Perceptions?},
  pdfauthor={Florian M. Farke, David G. Balash, Maximilian Golla, Adam J. Aviv},
  pdflang={en-US},
  pdfkeywords={privacy; targeted advertisement; online study; generic paradox; Google Ad~Settings},
  pdfdisplaydoctitle=true,
  bookmarksnumbered,
  colorlinks = true,
  pdfborder = {0 0 0},
  linkcolor={black!80!black},
  citecolor={red!70!black},
  urlcolor={blue!70!black},
  pdfstartview={FitH},
  breaklinks=true,
  hypertexnames=false
}

\newcommand*{\appref}[1]{\hyperref[#1]{Appendix~\ref*{#1}}}

\newcommand{\ie}{i.\,e.}
\newcommand{\eg}{e.\,g.}
\newcommand{\cf}{cf.\@\,}

\newcommand{\etal}{et~al.\@\,}

\newcommand{\adsettings}{Ad~Settings\xspace}
\newcommand{\gsearch}{Google~Search\xspace}
\newcommand{\yt}{YouTube\xspace}
\newcommand{\gmaps}{Google~Maps\xspace}

\newcolumntype{H}{>{\setbox0=\hbox\bgroup}c<{\egroup}@{}}
\newcolumntype{L}[1]{>{\hsize=#1\hsize\raggedright\arraybackslash}X}%
\newcolumntype{R}[1]{>{\hsize=#1\hsize\raggedleft\arraybackslash}X}%
\newcolumntype{C}[2]{>{\hsize=#1\hsize\columncolor{#2}\centering\arraybackslash}X}%

\newlist{compactitem}{itemize}{5}
\setlist[compactitem]{leftmargin=*, nosep}
\setlist[compactitem, 1]{label=\textbullet}
\setlist[compactitem, 2]{label=\textendash}
\setlist[compactitem, 3]{label=\textasteriskcentered}
\setlist[compactitem, 4]{label=\textperiodcentered}

\newlist{questions}{enumerate}{3}
\setlist[questions]{align=left, labelwidth=2em, labelsep=.5em, listparindent=0pt, itemindent=0pt, leftmargin=!, format=\bfseries}
\setlist[questions, 1]{labelindent=0pt, label=Q\arabic*, widest=99}
\setlist[questions, 2]{labelindent=-2.5em, label*=\_\Alph*, widest=26}
\setlist[questions, 3]{labelindent=-2.5em, label*=\_\roman*, widest=9}

\newlist{answers}{itemize}{1}
\setlist[answers]{leftmargin=*, nosep, align=left, label=$\bigcirc$}
\newlist{answers*}{itemize*}{1}
\setlist[answers*]{label=$\bigcirc$}

\begin{document}
\title[Connecting Online Activities to Advertising Inferences]{How Does Connecting Online Activities to Advertising~Inferences Impact Privacy Perceptions?}

\author{Florian M. Farke}
\orcid{0000-0001-7138-4978}
\affiliation{%
\institution{Ruhr University Bochum}
\city{}
\state{}
\country{}
}
\email{florian.farke@rub.de}

\author{David G. Balash}
\orcid{0000-0002-9730-9949}
\affiliation{%
\institution{University of Richmond}
\city{}
\state{}
\country{}
}
\email{david.balash@richmond.edu}

\author{Maximilian Golla}
\orcid{0000-0003-2204-2132}
\affiliation{%
\institution{CISPA Helmholtz Center for Information Security}
\city{}
\state{}
\country{}
}
\email{golla@cispa.de}

\author{Adam J. Aviv}
\orcid{0000-0002-3792-2485}
\affiliation{%
\institution{The George Washington University}
\city{}
\state{}
\country{}
}
\email{aaviv@gwu.edu}

\renewcommand{\shortauthors}{Farke et al.}

\begin{abstract}
Data dashboards are designed to help users manage data collected about them.
However, prior work showed that exposure to some dashboards, notably Google's My~Activity dashboard, results in significant \emph{decreases} in perceived concern and \emph{increases} in perceived benefit from data collection, contrary to expectations.
We theorize that this result is due to the fact that data dashboards currently do not sufficiently \enquote{connect the dots} of the \emph{data food chain}, that is, by connecting data collection with the use of that data.
To evaluate this, we designed a study where participants assigned advertising interest labels to their own real activities, effectively acting as a behavioral advertising engine to \enquote{connect the dots.}
When comparing pre- and post-labeling task responses, we find no significant difference in concern with Google's data collection practices, which indicates that participants' priors are maintained after more exposure to the data food chain (differing from prior work), suggesting that data dashboards that offer deeper perspectives of how data collection is used have potential.
However, these gains are offset when participants are exposed to their true interest labels inferred by Google. Concern for data collection dropped significantly as participants viewed Google's labeling as generic compared to their own more specific labeling.
This presents a possible new paradox that must be overcome when designing data dashboards, the \emph{generic paradox}, which occurs when users misalign individual, generic inferences from collected data as benign compared to the totality and specificity of many generic inferences made about them.
\end{abstract}


\maketitle

\section{Introduction}\label{sec:intro}

Google provides multiple dashboards for customers to review and manage activity data collection and advertising personalization.
The \emph{My~Activity} interface provides a canonical list of all activities linked to a Google account, \eg, web searches, watched \yt videos, or map interactions.
The \emph{\adsettings} dashboard allowed consumers to review the \emph{ad interests} inferred or assigned based on these activities (following the completion of our study, the service was revised to \emph{My~Ad~Center}~\cite{google-22-my-ad-center}).
Prior work on dashboards \cite{zimmermann-14-privacy-dashboards, fischer-huebner-16-priv-dashboard, raschke-17-usable-pri-dashboard, herder-20-privacy-dashboard} suggests that they can increase understanding of data collection and online advertising~\cite{ur-12-creepy, rader-14-awareness-google, chanchary-15-percep-tracking, tschantz-18-accu-google-ad, dolin-18-dd-inferences, weinshel-19-places-you-been, rader-20-inferences-vs-perceptions, wei-20-twitter-knows}.

However, our previous work~\cite{farke-21-privacy-dashboads} indicates that dashboards can have a surprising, inverse effect. Exposure to Google's My~Activity dashboard led to a significant \emph{decrease} in perceived concern and an \emph{increase} in the perceived benefit of data collection.
At the time, we suggested that the disconnect between what data is collected and how that data may be used could be responsible for the dramatic shift in concern and benefit before and after viewing the dashboard.
In this work, we seek to explore this disconnect between data collection and usage.
We theorize that dashboards, which remain disconnected from the \emph{data food chain}~\cite{nissenbaum-19-data-food-chain} by not demonstrating to users how their activities are used, are a key contributor to this observation.
Nissenbaum's data food chain metaphor is a conceptual framework that illustrates the flow of personal information through a hierarchy of stages of processing and usage, from the initial collection of raw data to the generation of actionable insights.

We evaluate this theory to determine if \enquote{connecting the dots} in the data food chain, \ie, connecting activities to advertising interests, stabilizes perceived benefit of and concern for online data collection.
We conducted an interactive online study with \num{170} participants who installed a browser extension to review a selection of their actual Google activities, such as those from Search, YouTube, and Maps, across recent to 18-month-old interactions. Participants then assigned ad interests they believed Google learned from those activities, effectively \enquote{reenacting} the task of Google's behavioral advertising algorithms. Using a pre- and post-study design, we questioned participants about their concerns and perceived benefits of Google's data collection before and after they completed assigning advertising labels to their online activities. This approach provided insight into changes in attitudes towards data privacy due to direct engagement with the data that informs ad personalization.
Through this methodology, we sought to answer the following research questions:
\begin{enumerate}[leftmargin=*,label=RQ\arabic*,format=\bfseries,topsep=5pt,itemsep=5pt,partopsep=0pt,parsep=0pt,labelwidth=2pt]
    \item\label{RQ:understanding-inferencing} \emph{[Understanding of Ad Interests] How well do people understand inferred ad interests based on assigning interests to their real activities?}
    
    \item\label{RQ:matching-effect} \emph{[Impact on Benefit/Concern] How do people's concerns and impressions about the benefits of Google's data collection change after assigning interests to their real activities?}
    
    \item\label{RQ:reactions-interests} \emph{[Reactions to Ad Interests] How do people react to the actual interests Google assigns to them?}
\end{enumerate}

In answering \ref{RQ:understanding-inferencing}, participants assigned interest labels very differently compared to their \adsettings page: only \num{21}\% of assigned interests occurred in \adsettings.
Those not in \adsettings{} were often overly specific interests (\eg, a specific sports team), while \adsettings interests were more general (\eg, \enquote{sports}).
This expected specificity, which contrasts the generic labels on the \adsettings{} page, greatly impacts perceptions of concern and benefit when answering \ref{RQ:matching-effect}.
Participants were asked Likert questions about Google's data collection before and after the labeling task, and similar to our previous work~\cite{farke-21-privacy-dashboads}, there was a significant increase (small effect) in benefit following the labeling task but no significant difference with respect to concern, differing from prior work.
This difference suggests that exposure to the data food chain in data dashboards can maintain priors, at least with respect to the participants' concerns about data collection.
However, we find that there is a significant decrease (moderate effect) after participants viewed their actual Google-assigned advertising interests, while the perceived benefit from Google's data collection remains unchanged.
In answering \ref{RQ:reactions-interests}, most participants found these interests at least somewhat accurate.
Many felt they were too generic to be a concern, which is in line with the work by Radar~\etal~\cite{rader-20-inferences-vs-perceptions}.
These results provide insights that may be applicable to various advertisement preference managers, including those offered by Facebook~\cite{facebook-23-ad-preferences}, X (formerly Twitter)~\cite{x-23-ad-preferences}, Microsoft~\cite{microsoft-23-ad-preferences}, LinkedIn~\cite{linkedin-23-ad-preferences}, and Amazon~\cite{amazon-23-ad-preferences}, whose platforms similarly aim to give users control over their collected activity data and the personalized ads based thereon.

While there are opportunities for data dashboards to better connect to the data food chain, there may also be a new paradox to overcome. We describe a new, potentially observed one as the \emph{generic paradox}, where the perceived generality of data collection and its usage counteracts the combined specificity and privacy impact of multiple generic inferences. For data dashboards to be most effective in assisting users, they should both provide connections to the data food chain and combat the generic paradox, such as offering examples of how seemingly generic combinations of labels are actually quite specific.
For example, while a single ad interest presented to the user in the Google \adsettings interface is generic, Google and advertisers could aggregate these interests and infer a more detailed picture of a user than any single interest can on its own.

\section{Related Work and Background }\label{sec:related-work}

\paragraph{Related Work}
There is a rich body of related work, but we mostly focus on two specific areas. First is research that investigates users' understandings of the online behavioral advertising ecosystem~\cite{rader-20-inferences-vs-perceptions, warshaw-16-intuitions, barbosa-21-who-am-i, weinshel-19-places-you-been}, e.g., by conducting a survey of the topic or letting users interact with their own labels. Our study builds upon this line of work as we are also concerned with how users perceive this ecosystem; however, we apply a different method. In particular, we ask participants to not only observe the labels applied to them by advertisers but also to imagine generating said labels.
The second major area focuses on measurements within the advertising ecosystem~\cite{sabir-22-facebook-activity, wills-12-understanding-what-they, datta-15-automated-experiments, andreou-19-measuring-facebook-advertising, reitinger-23-ad-settings}, such as its accuracy and the frequency in which the inferences are updated or individualized. Our work builds on this line of research as well by observing directly how users expect labels to be applied compared to how they are actually applied as a proxy measurement of the accuracy of the labels.

Most of all, our research is motivated by the findings of our previous study~\cite{farke-21-privacy-dashboads}. There we conducted a study with \num{153}~participants to examine the impact of Google's My~Activity dashboard on participants' concerns and benefits of Google's data collection.
Surprisingly, interacting with the dashboard led to a decrease in concern and an increase in perceived benefits, and we discussed the possibility that such an effect could be mitigated by providing more context for how the data is used in the dashboard. Our work builds on that conjecture through our survey methodology, namely the labeling task as the primary task in a pre-post study design.

In the rest of this section, we describe in more detail the prior work we build upon. Rader \etal~\cite{rader-20-inferences-vs-perceptions} investigated people's reactions to Google's and Facebook's ad interests and found that understandings are bounded by past online behaviors and self-perceptions of interests.
To investigate inference literacy through interviews, Warshaw \etal~\cite{warshaw-16-intuitions} found that few believed companies can make the type of deep personal inferences that are nowadays common and that inferences are based largely on directly gathered demographic data or via straightforward processing of online behavioral data.
Barbosa \etal~\cite{barbosa-21-who-am-i} built a tool to help individuals engage with automatically constructed profiles by displaying in-the-moment notifications of the profile construction.
They found that this increased visibility and understanding of how particular actions can lead to specific ads.
To understand how visualizing inference-level information about online tracking impacts people's knowledge, perceptions, and attitudes, Weinshel \etal~\cite{weinshel-19-places-you-been} built a browser extension to conduct a longitudinal field study, finding that after participants viewed visualized examples, they had a more accurate perception of tracking and were more likely to take action. 
Sabir \etal~\cite{sabir-22-facebook-activity} explored how Facebook generates ad~interests and found that even minor activity, such as scrolling through a page, can lead to an inference. But interests can often be inaccurate with explanations that are too generalized and/or misleading. 

Datta \etal~\cite{datta-15-automated-experiments} created \enquote{AdFisher} to explore how individual behaviors impact Google's ads and inferences.
Rather than providing transparency, \adsettings was opaque about some features of a person's profile, and visiting certain websites could change the ads but not the reported inferences.
Wills \& Tatar~\cite{wills-12-understanding-what-they} used controlled browsing and also found cases in which ads were shown to people but did not appear as categories.
To determine what advertisers infer about individuals and their accuracy, Bashir \etal~\cite{bashir-19-quantity-vs-quality} studied four ad~preference managers from Google, Facebook, Oracle~BlueKai, and Nielsen~eXelate by gathering full interest profiles and asking participants if they were actually interested in these topics. 
They found that participants were strongly interested in only $27\%$ of the interests in their profiles and that browsing history only explains a small percentage of interests.
Andreou \etal~\cite{andreou-19-measuring-facebook-advertising} measured the Facebook advertising ecosystem and found that the median number of interests inferred for a person is $310$ and that a significant number of advertisers employ targeting strategies that could be either invasive or opaque.
Finally, recent work by Reitinger \etal~\cite{reitinger-23-ad-settings} conducted a longitudinal measurement study showing that the inferred interests on Google \adsettings are frequently updated, accurate, and often individualized. Confirming their findings, we discuss the dangers of combining inferences in Section~\ref{sec:discussion}.

\paragraph{Google's Advertising Interests}\label{sec:background}
Google offers targeted advertising as a tool for advertisers to display their ads to selected audiences based on a profile of \enquote{interests,} \enquote{purchase intentions,} or \enquote{behaviors.}
Google determines these profiles based on user interactions with Google's \enquote{free} services, \eg, Chrome, \gsearch, \gmaps, and \yt, where, for example, watching \emph{Star Wars} leads to the interest \emph{Science Fiction \& Fantasy Films} to appear in a profile.
Users can also view/delete individual interests, limit certain categories (\ie, alcohol and gambling), or completely disable ad~personalization on the \adsettings page (\href{https://adssettings.google.com}{adssettings.google.com}) (revised to \emph{My~Ad~Center} in November 2022~\cite{google-22-my-ad-center}).

At the time of the study, Google described five different interest types:
\begin{enumerate*}[label=(\roman*)]
    \item \emph{Demographic} interests (age, gender, and language) of self-reported data,
    \item \emph{Advertiser} interests (\eg, \enquote{Samsung Electronics}), and
    \item \emph{Aggregated} interests (\eg, Homeownership Status) by determining activities \enquote{similar to people who've told Google they're in this category.}
By far, the biggest number of inferences are 
\item \emph{Activity} interests that are directly inferred from the activities associated with a Google account, and also a few 
\item \emph{YTC} interests, likely based on \yt comments or \enquote{video interactions.}
\end{enumerate*}
Google describes around \num{2453}~interests that originate from their AdWords \emph{vertical} list~\cite{google-21-adwords-verticals} that starts with 26~topics ranging from \enquote{Arts \& Entertainment} to \enquote{Travel \& Transportation.}
The remaining \num{2427} labels are subcategories (or further nested) of these topics, with up to eight levels.
We use this list of interests for participants to auto-fill when labeling their own activities.
A list of example interests can be found in \autoref{tab:example-interests} in \appref{app:tables}.
Other services like Facebook, LinkedIn, or Microsoft offer similar functions to control ad~inferences.

\section{Method, Recruitment, and Limitations}\label{sec:study}

\begin{figure*}[t]
  \centering
  \includegraphics[width=1.0\linewidth]{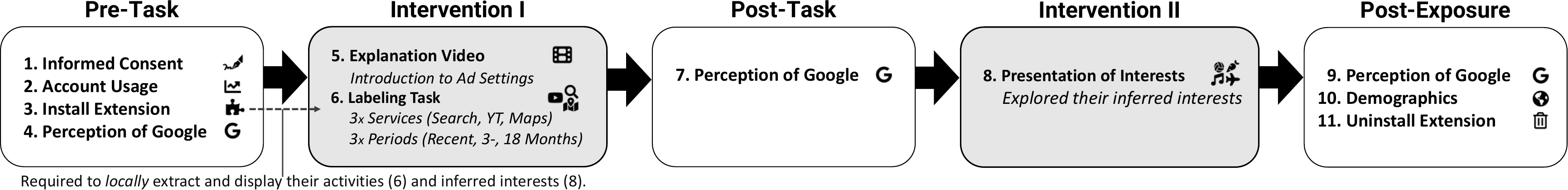}
  \caption[Study Protocol]{\label{fig:study-design}
    The study consisted of five parts.
    During the first intervention, participants labeled recent, 3-, and 18~months-old activities from Search, YouTube, and Maps and during the second intervention, they explored their \adsettings interests.
  }
  
\end{figure*}

The study was designed around a custom online survey and browser extensions, in which participants interacted with their real Google account activities to apply interest labels, the \emph{labeling task} (described below).
At the time of the study, Google's My~Activity and \adsettings dashboards only show the result, either the activity or the interest, but not how activities connect to interests. 
In conversation with our prior work~\cite{farke-21-privacy-dashboads}, this study is designed around a labeling task to evaluate how reflective thinking~\cite{fleck-10-reflecting-reflection} may alter the mental models of the data collection and assist users in understanding how their data may be used, which  could lead to changes in attitude about technology~\cite{bateman-12-search-dashboard}.
Below, we describe the study procedure and analysis in more detail.

\paragraph{Study Procedure}\label{sec:study-procedure}
We performed six pilots with co-workers to refine the protocol, and we also performed test runs ($n=10$) on Prolific~\cite{prolific-22-user-study}, fixing an issue with timestamps at the start and end of the survey.
The final procedure is outlined in Figure~\ref{fig:study-design} and detailed below. The complete survey can be found in \appref{app:survey}.

\begin{enumerate}[itemsep=4pt]

    \item \textit{Informed Consent:}
     Participants consented to the described procedure, risks, benefits, and right to withdraw.

    \item \textit{Google Account Usage (Questions \ref{app:survey:s1}--\ref{app:survey:s5}):}
    Intended as warm-up questions, participants were asked about their Google account age, usage, and importance.

    \item \textit{\enquote{Survey Assistant} Browser Extension and Google Sign-In:}
    Participants were directed to install the browser extension, with descriptions of its functionality and removal after the study.
    After this, participants signed in to their primary Google account using third-party sign-on and were informed that no details beyond their My~Activity and \adsettings pages would be accessed.

    \item \textit{Pre-Task Perceptions (Questions \ref{app:survey:q1}--\ref{app:survey:q3}):}
    Participants answered pre-task questions regarding awareness~(\ref{app:survey:q1}), concern~(\ref{app:survey:q2}), and benefit~(\ref{app:survey:q3}) of Google's data collection practices.
         
    \item \textit{Explanation of Google \adsettings (Video):}
    Participants watch\-ed a short video~\cite{google-20-ad-settings-video} by Google explaining \adsettings.

    \item \textit{Labeling Tasks:}
    Participants were presented with their past activities parsed from My~Activity for three services, Google Search, \yt, and Maps, and time periods, recent (\ie, last seven days), 3-months old, and 18-months old~\cite{pierce-20-google-3-and-18-months}. We randomly selected 100 activities that met these requirements and randomly displayed 9 (three services $\times$ three time periods).
    Participants were asked to assign ad interests they believe Google learns from the shown activities.
    Participants were allowed to skip activities (when they considered the activity boring, generic, or sensitive), replacing it with another random activity of the same criteria. If all activities in a category were skipped, the entire portion was skipped. 
    Only participant-labeled activities are included in the analysis.
    
    \item \textit{Post-Task Perceptions (Questions \ref{app:survey:q9}, \ref{app:survey:q10}):}
    Participants were asked about their concern (\ref{app:survey:q9}) and benefit (\ref{app:survey:q10}).

    \item \textit{Exposure to Inferred Interests:}
    Participants were then presented their real interests from Google, which must be viewed for 30~seconds before continuing. 
    This part and the next were skipped when participants turned off ad~personalization.

    \item \textit{Post-Exposure to Inferred Interests (Questions \ref{app:survey:q11}--\ref{app:survey:q13}):}
    Participants were asked again about concern~(\ref{app:survey:q11}) and benefit~(\ref{app:survey:q12}) and if they perceived their interests as accurate~(\ref{app:survey:q13}).

    \item \textit{Demographics (Questions \ref{app:survey:d1}--\ref{app:survey:d10}):}
    Participants (optionally) provided demographic data.
    
    \item \textit{Browser Extension Removal \& Feedback:}
    Participants gave feedback optionally and removed the extension.
\end{enumerate}

\paragraph{Statistical Tests and Regression Analysis}
We performed six \emph{Wil\-coxon signed-rank tests} for repeated measurements on the Likert responses to the questions on concern \ref{app:survey:q2}, \ref{app:survey:q9}, and \ref{app:survey:q11}, as well as the questions on frequency of benefit \ref{app:survey:q3}, \ref{app:survey:q10}, and \ref{app:survey:q12}.
To correct the p-values, we used \emph{Holm's method} for overlapping measures. These tests assume that the exposures, either the labeling task or viewing the interests, affected changes in concern or benefit.

To explore if there are other factors, we conducted two \emph{binomial logistic regressions} that included demographics, labeling accuracy, and other factors to determine what else might influence participants to decrease their concern rating or increase their benefit rating, respectively. 
Finally, we calculated a \emph{binomial logistic regression mixed-effect model} to explore whether participants were more likely to choose labels that Google also inferred (\ie, were on their Ad Settings page) for activities of a specific service or time frame.

\paragraph{Qualitative Coding}
We analyzed the free-response questions on concern \ref{app:survey:q2a}, \ref{app:survey:q9a}, and \ref{app:survey:q11a}, frequency of benefit \ref{app:survey:q3a}, \ref{app:survey:q10a}, and \ref{app:survey:q12a}, and accuracy of the \adsettings page \ref{app:survey:q13a} using \emph{qualitative open coding}.
One author bootstrapped the process using the codebook published in the extended version of our previous work~\cite{farke-21-privacy-dashboads} by coding the entire set of responses.
Whenever they identified missing themes, new codes were added.
A new codebook was crafted for the accuracy question using the same method.
To validate it, a secondary coder conducted several rounds of coding on a $20\%$ sub-sample of each  free-response question.
Every round of coding resulted in updating the codebook.
We repeated this process until we reached an inter-coder agreement (\emph{Cohen's} $\kappa > 0.7$, $mean = 0.74$, $sd = 0.02$).

Since we were especially interested in why people change their concern/benefit rating in the different parts of the survey, we did another coding with the subset of participants that changed their rating.
Here, the coders identify reasons for increasing/decreasing concern/benefit, and this process was repeated together with a second coder for the question pairs \ref{app:survey:q2a} and \ref{app:survey:q9a}, as well as \ref{app:survey:q3a} and \ref{app:survey:q10a}.

\paragraph{Ethical Considerations}
The institutional review board~(IRB) of the George Washington University approved our study under following number \emph{NCR213523}.
All participants were informed and consented to the study.
As our browser extension had access to the participants' Google accounts, we were especially cautious not to collect any data participants would not want to share.
At no time did the extension nor the researchers have access to the participants' Google password or any other Google account data besides My~Activity and \adsettings.
Participants could also decide in each labeling task whether they consider the presented activity too sensitive and thus exclude it from the analysis.
We also carefully reviewed the final dataset for personally identifiable information (PII) and removed anything we considered problematic.

\paragraph{Recruitment and Demographics}
Of the initially recruited \num{318}~people residing in the U.S. via Prolific~\cite{prolific-22-user-study}, \num{235} completed the study in May 2022.
We do not consider the dropout rate a systematic issue with the participant pool, as the population of Prolific generally yields representative results for questions about
user perceptions and experiences~\cite{redmiles-19-mturk-generalization, tang-22-external-validity}.
We rather explain the number of dropouts as follows:
\num{41}~people did not consent to participate in the study (either they did not want to install the browser extension or were using the wrong browser/device), \num{19} more chose not to log into their Google account, and for \num{9}, the browser extension failed to fetch data.
Four additional people completed the study but failed to submit for credit on Prolific, and four withdrew mid-study.
Only the remaining six people dropped out for unknown reasons.

All \num{235}~participants who completed the study were compensated \SI{5.00}[\dollar]{\USD} with a  medium completion time of \num{16} to \num{17}~minutes.
As the study consisted of interactive tasks (\eg, authenticating with Google, installing an extension, and answering open-text responses), we did not use attention check questions due to the ability to assess data quality directly.
In reviewing responses, we observed only four participants who were eventually excluded due to inconsistencies in their data.
Not all participants had sufficient activities for the labeling tasks, and only participants who completed at least three tasks were considered in the analysis, excluding another \num{61}~participants.
These eligibility checks resulted in \num{170}~participants being included in the final analysis.
\autoref{tab:demographics} in \appref{app:tables} shows the demographics of the \num{235}~participants who completed the study and the \num{170}~participants included in the analysis. 

Of these \num{170} participants, \num{133}
had Google ad~personalization enabled with a total of \num{22661} different interests, 
with a median of \num{180}~interests ($SD = 73$) per participant.
Most interests were activity-based ($n = \num{20175}$, representing \num{1230} distinct interests).
The remaining \num{2486} interests included: \num{1258} advertiser interests from \num{618} different advertisers; \num{745} aggregated interests with \num{38} different interest combinations across 7~categories; \num{393} demographic interests with \num{17} distinct interests across 3~categories); and \num{90} YTC interests, of which \num{85} were distinct.
The top~10 interests, which are all activity based, can be found in \autoref{tab:top-interests} in \appref{app:tables}.

We did not ask the \num{37} participants why they disabled ad~personalization. Privacy concerns may be one possible explanation, but we did not find any significant difference in the initial level of concern between both groups.

\paragraph{Limitations}
Due to the restrictions of our recruitment, our sample is not fully representative in multiple dimensions. 
The study's methodology inherently introduces a degree of self-selection bias, particularly in the requirement for participants to install a browser extension and share their Google account data. This requirement may deter more privacy-conscious individuals from participating, as they may be less willing to grant access to their online activities and install third-party software.
Despite this, we argue that there are still significant and relevant takeaways, particularly as recent work suggests that online samples can provide insights compared to the general population~\cite{redmiles-19-mturk-generalization, tang-22-external-validity}.
Notably, Tang et al.~\cite{tang-22-external-validity} showed that gender-balanced data collected through Prolific for online privacy and security surveys is generally representative when questioning participants about their perceptions and experiences. While only studying an U.S.-based sample is a limitation, it was chosen for better comparability~\cite{farke-21-privacy-dashboads}.

There are also limitations in data balancing as a number of participants did not complete the same number of labeling tasks, either because they skipped activities or did not have enough activities. We argue that the exposures to activities and advertising interests are still sufficient to measure pre- and post-effects as well as to perform exploratory regression analysis.
These limitations may, however, affect results regarding the difference in labeling and perceptions thereof, and we include clarifications in those cases.
There was also a reasonable fraction ($n = 37$) of participants who disabled ad~personalization, and as a result, some results are based on a smaller, but still meaningful, share of \num{133}~participants.

Finally, as our sample size, including the exclusions, is relatively small, there are limitations regarding the conclusions that can be drawn from failed significance testing.
We conducted power analyses and determined that the Wilcoxon sign-ranked tests comparing pre- and post-exposure should be sufficiently powered with \num{170} to identify medium effects ($w=0.3$).
For the regression analyses with 12~predictors, the power analyses revealed that the regression models are also sufficient for identifying medium effects ($f=0.15$).
While this is reasonable for identifying trends, a much larger sample would be needed to rule out other influential factors that might have small effect sizes conclusively.

\section{Results}\label{sec:results}
This section is structured along our three research questions.
We first present participants' understanding of inferred ad interests by describing how they labeled their online activities.
Secondly, we show the impact of letting participants connect their online activities and interest labels on the perceived concern and benefit of Google's data collection.
Finally, we report users' perceptions of the interest labels that Google actually applied to their profile.

\subsection{\ref{RQ:understanding-inferencing}: Understanding of Ad Interests}

\paragraph{Labeling Task}
Participants were presented with nine activities from a list of up to 100 randomly selected activities from their My~Activity page that matched the following criteria.
Activities were selected both temporally and based on services: recent, three months, or 18~months old~\cite{pierce-20-google-3-and-18-months} and \gsearch, \yt, or \gmaps.
Participants could skip activities if they found them \enquote{too boring,} \enquote{too generic,} or \enquote{too sensitive.}
When skipped, the activity was replaced with another one from the list of selected activities.
Most participants, \ie, \num{119} of the \num{170} participants, skipped at least one activity with a median of seven activities, and \num{27} participants skipped enough activities of one criterion, only labeling 8 activities instead of 9.
Only two participants skipped two of the nine labeling tasks. 

\begin{figure}[t]
  \centering
  \includegraphics[width=\columnwidth]{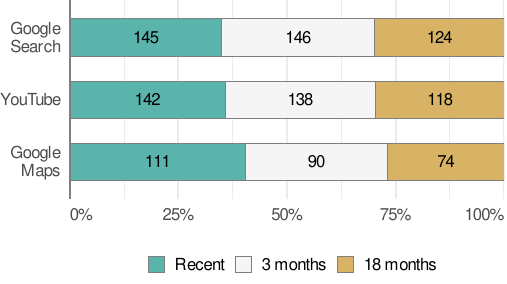}
  \caption[Number of completed labeling tasks]{\label{fig:labeling-task-frequency}
    Number of activities labeled by participants during the labeling tasks grouped by service and age of the activity.
  }
\end{figure}

Overall, \num{31} participants completed nine and \num{30} participants eight of the labeling tasks, respectively.
Seven labeling tasks were completed by \num{23}~participants, six by \num{33}, and five by \num{18}~participants.
The remaining 35~participants completed four ($n = 15$) or three ($n = 20$) labeling tasks, most often because there were not enough activities in those participants' profiles to match the criteria.
We include these \num{170}~participants who completed at least three labeling tasks in the analysis below. \autoref{fig:labeling-task-frequency} shows a breakdown of the completed labeling tasks per service and time frame.

\begin{figure}[t]
  \centering
  \includegraphics[width=\columnwidth]{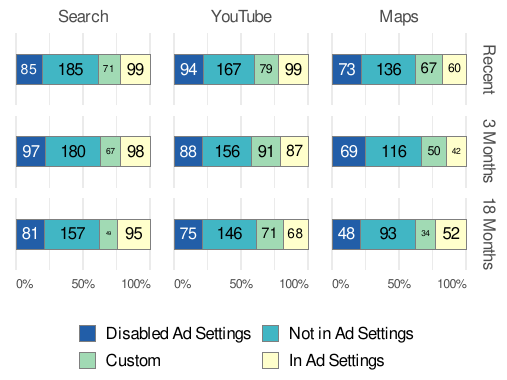}
  \caption[Interest labels in labeling task]{\label{fig:interst-labels-labeling-tasks}
    Number of labels applied during the labeling tasks that are present in participants' \adsettings page (\emph{In \adsettings} vs. \emph{Not in \adsettings} series) grouped by service and age.
    The \emph{Disable \adsettings} series comprise the labels of participants with an empty \adsettings page and \emph{custom} accounts for participants' labels not in the list of \emph{verticals}~\cite{google-21-adwords-verticals}.
  }
\end{figure}

\paragraph{Participants' Labels Present in \adsettings Page}
Participants labeled activities with advertising interests by typing into a text field that suggested \emph{verticals} from AdWords~\cite{google-21-adwords-verticals}.
Additionally, they could create and add custom labels.
We compared the labels applied by the participants with the list of interests from their \adsettings page. 
\autoref{fig:interst-labels-labeling-tasks} visualizes the difference between the participants' self-applied labels and Google's applied interest labels.
In total, the participants applied \num{3325} labels across all completed labeling tasks.

Participants often created a custom label even when a matching vertical was available. We developed multiple processes to match custom labels to verticals.
First, we performed simple text transformations, \eg, adjusting the capitalization and striping white spaces.
Next, we combined any custom labels that were synonyms (\eg, \enquote{Kids} and \enquote{Children}), singular/plural writings, and corrected typos.
We then split the custom labels into individual words, filtering out any \emph{stop words} (\eg, function words like \enquote{and} or \enquote{to}), and removed all remaining words with less than three letters.
The remaining words were again compared to AdWords verticals and replaced with a vertical if a match existed.

We sought agreement between two researchers to verify replacements.
One researcher reviewed if one of the replacement verticals was suitable (\eg, \enquote{Computer~\&~Video Games} for \enquote{Video~Games}) and aligned the decision with that of the other researcher.
The custom label was not replaced when no agreement was reached.
This process reduced the number of custom labels from \num{689} to \num{584} distinct labels.
Finally, we performed a qualitative coding of the remaining custom labels as \emph{specific} and \emph{generic} labels.

Most of the selected labels did not occur on the participants' \adsettings page (on median, only \num{5} out of \num{21}).
As shown in \autoref{fig:interst-labels-labeling-tasks}, these label differences occurred across services and time frames, and we sought to explore what factors may lead to more concurring labels. To do so, we calculated a generalized logistic regression with mixed effects (see \autoref{tab:labeling-task-regression}), with an outcome variable of whether the labels applied by the participants are on their \adsettings page.
We included the three services and time frames, with interactions as input variables.
Since participants performed multiple labelings (\ie, repeated measures), the participant's~ID is a random effect, and \yt and \num{3}~months are the model baselines.

We find that when selecting Google Maps, participants are less likely to choose labels showing up on their \adsettings page, except potentially when considering a Maps~activity that is 18~months or older (trending towards significance).
However, the model's \emph{Aldrich-Nelson} pseudo-$R^2$ is \num{0.0069}, which is low, suggesting that these factors do not explain the variance in labeling difference well. 

\begin{table}[t]
\centering
\caption[Labeling Task Regression Model]{\label{tab:labeling-task-regression}
Binomial logistic regression model with mixed-effect to predict whether participants apply interests labels that are on their \adsettings page.
We are using a mixed-effect model as the same participant labeled multiple activities resulting in a mix of within-subject and between-subject measurements.
The Aldrich-Nelson pseudo $R^2$ of the model is \num{0.0069}.
}%

\footnotesize
\begin{tabularx}{\columnwidth}{@{}X@{\extracolsep{\fill}}rrrrr@{~}l@{}}
\toprule
\textbf{Factor} &
\multicolumn{1}{@{}c@{}}{\textbf{Est.}} &
\multicolumn{1}{@{}c@{}}{\textbf{OR}} &
\multicolumn{1}{@{}c@{}}{\textbf{Error}} &
\multicolumn{1}{@{}c@{}}{\textbf{z}} &
    \multicolumn{2}{@{}c@{}}{\textbf{Pr(\textgreater{}\textbar{}z\textbar{})}} \\
\midrule
(Intercept)               & \textbf{-1.11} & \textbf{0.33} & \textbf{0.15}  & \textbf{-7.63} & \textbf{\textless{}0.001} & ***     \\
\gsearch                  & 0.05  & 1.06 & 0.18  & 0.30  & 0.767            &         \\
\gmaps                    & \textbf{-0.42} & \textbf{0.65} & \textbf{0.23}  & \textbf{-1.88} & \textbf{0.059}            & .       \\
Recent                    & 0.11  & 1.12 & 0.18  & 0.61  & 0.541            &         \\
18 months                 & -0.16 & 0.85 & 0.20  & -0.84 & 0.402            &         \\
\midrule
\textbf{Interaction} &
\multicolumn{1}{@{}c@{}}{\textbf{Est.}} &
\multicolumn{1}{@{}c@{}}{\textbf{OR}} &
\multicolumn{1}{@{}c@{}}{\textbf{Error}} &
\multicolumn{1}{@{}c@{}}{\textbf{z}} &
    \multicolumn{2}{@{}c@{}}{\textbf{Pr(\textgreater{}\textbar{}z\textbar{})}} \\
\midrule
Search $\times$ Recent    & -0.13 & 0.87 & 0.25  & -0.53 & 0.595            &         \\
Maps   $\times$ Recent    & 0.14  & 1.15 & 0.30  & 0.47  & 0.639            &         \\
Search $\times$ 18~months & 0.25  & 1.29 & 0.27  & 0.96  & 0.339            &         \\
Maps   $\times$ 18~months & \textbf{0.60}  & \textbf{1.83} & \textbf{0.32}  & \textbf{1.91}  & \textbf{0.056}            & $\cdot$ \\
\bottomrule
\end{tabularx}
\footnotesize
\textbf{Signif. codes:} $\text{***}~\widehat{=} < 0.001$; $\text{**}~\widehat{=} <0.01$; $\text{*}~\widehat{=} <0.05$; $\cdot~\widehat{=} <0.1$
\end{table}

\paragraph{How Participants Labeled Their Activities}
To observe the differences between the labels the participants and Google applied, we must exclude \num{710} of the \num{3325} labels applied by participants that had disabled \adsettings.
Out of the remaining \num{2615} applied labels, \num{700} are on the participants' \adsettings page, \num{1336} did not, and \num{579} were custom labels created by the participants.
There are two potential reasons for this:
\begin{enumerate*}[label=(\roman*)]
    \item the activity was unusual for the participant, or
    \item participants apply interests differently than Google.
\end{enumerate*}
We reviewed the non-matching labels and found that there are two  distinct groups of labels applied by participants not appearing in Google's labeling:
\begin{enumerate*}[label=(\roman*)]
    \item Too generic (\eg, \emph{Comedy}, \emph{Maps}, \emph{Images}, \emph{YouTube}, \emph{knowledge}); or
    
    \item overly specific (\eg, \emph{Minecraft}, \emph{Disney+}, \emph{Taylor Swift Red}).
    
Other reasons include the use of 
    \item banned or sensitive ad categories (\eg, \emph{Christianity}, \emph{pornography}, \emph{COVID-19}, \emph{brewing}, \emph{depression}) or
    
    \item context-dependent knowledge that cannot be inferred from the activity alone (\eg, \emph{hobbies}, \emph{ideas}).
\end{enumerate*}

Among the three services, Google Maps activities were the most difficult to label, with participants often applying labels like \emph{recently visited}, \emph{places}, \emph{street}, or \emph{driving directions}.
We also found that \num{231} out of the \num{3325} labels appeared in the activity description, \ie, the label \enquote{\emph{Reddit}} was used for the activity \enquote{\emph{Searched for: Can other people on Reddit see your followers}.}
When we split multi-word labels and removed stop words,  we even found matches for \num{490} out of \num{3325} labels, \eg, for the custom label \enquote{\emph{Rice Dishs[sic]}} in the \enquote{\emph{Searched for: Dye rice}.}

\subsection*{\emph{Summary of RQ1 [Understanding of Ad Interests]}}
We find that participants apply interest labels differently than Google, as only about one quarter of the applied labels occur on the participants' \adsettings pages. Google Maps activities are especially difficult to label.
Non-matching labels were either too generic~(\enquote{Maps}) or overly specific~(\enquote{Sims~4}).

\subsection{\ref{RQ:matching-effect}: Impact on Benefit and Concern}

\begin{figure*}[thbp]
  \begin{subfigure}[t]{\columnwidth}
    \centering
    \includegraphics[width=\columnwidth]{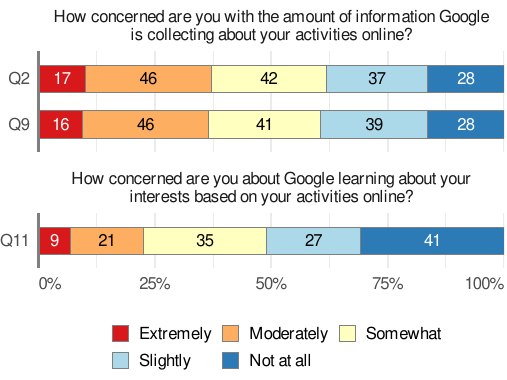}
    \caption[Level of concern throughout the survey]{\protect\label{fig:q2-q9-q11-bar}
      Level of concern throughout the survey.
    }
  \end{subfigure}
  \hfill
  \begin{subfigure}[t]{\columnwidth}
    \centering
    \includegraphics[width=\columnwidth]{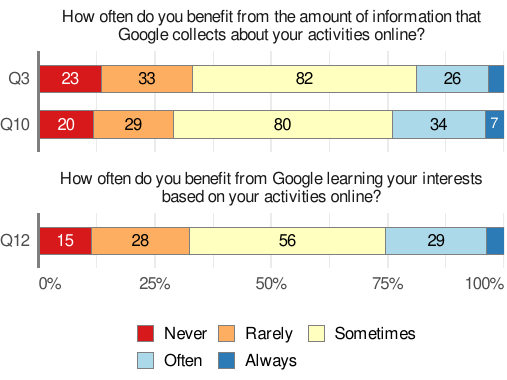}
    \caption[Frequency of benefit throughout the survey]{\protect\label{fig:q3-q10-q12-bar}
      Frequency of benefit throughout the survey.
    }
  \end{subfigure}
  \caption[Proportions of the level of concern and frequency of benefit]{\protect\label{fig:concern-benefit-bar}
    Participants' (a) level of \emph{concern} (\ref{app:survey:q2}, \ref{app:survey:q9}, \& \ref{app:survey:q11}) and
    (b) frequency of \emph{benefit} (\ref{app:survey:q3}, \ref{app:survey:q10}, \& \ref{app:survey:q12})
    at the beginning of the survey (Pre-Task), after the labeling tasks (Post-Task), and after exposure to the inferred interests (Post-Exposure).
  }
\end{figure*}

In this section, we focus on participants' perceived concern about and benefit from Google's data collection.
We ask the same Likert and open-response questions before and after the labeling task.
A shift in the responses may indicate a change in perspective due to the labeling task.
We provide a complete analysis of those responses and qualitative feedback regarding concerns/benefits before and after the task.

\paragraph{Initial Perceptions Concerns}
In \autoref{fig:q2-q9-q11-bar} (\emph{top bar}), the results of \ref{app:survey:q2} are presented regarding the concern for Google's data collection.
There is a near-even split between participants who were \emph{moderately} or \emph{extremely} concerned ($n = \num{63}$) and those who were \emph{slightly} or \emph{not at all} concerned ($n = \num{65}$).
Another \num{42} participants were \emph{somewhat} concerned.
These heterogeneous results indicate that participants have diverse perceptions of Google's data collection.

When participants were asked to explain their concerns (\ref{app:survey:q2a}), \num{86} out of \num{170} mentioned personal privacy.
Among these participants, \num{22} mentioned concerns about Google selling their data, such as P174, who reported being \emph{slightly concerned} and said, \enquote{I do not like that they can make tons of money selling my data but I don't see any of that money.}
Participants may be confounding selling targeted advertisements with selling personal information.
As illustrated by P78 (\emph{moderately concerned}), \enquote{I'm sure they are selling user info, if not directly then by selling targeted ads.} 

Sharing or having others view their personal information was a concern of \num{17} participants, \eg, P157 (\emph{extremely concerned}) responded, \enquote{I do not like strangers knowing my personal information and/or details of all my online activities.} 
Another concern mentioned by \num{16} participants was the amount of information collected.
As P153 (\emph{moderately concerned}) explained, \enquote{Google collects so much information that I'm concerned I don't really have a say in what they are collecting.} 
\num{11} participants were worried about third parties accessing their data, like P115 (\emph{moderately concerned}) who wrote:
\enquote{They have access, which I have granted them, to a lot of my private information. This information may end up with third parties.}

Further concerns include information being used to track people across websites and services were raised by \num{10} participants, such as when P133 (\emph{extremely concerned}) said, \enquote{I want my browsing experience to be private and not tracked.}
Moreover, \num{9} participants had concerns about the sensitive nature of some of the information, \eg, P142 (\emph{extremely concerned}) shared, \enquote{It concerns me that some of that information that is shared might be sensitive and something I do not want companies to know.}

For \num{29} participants, the unknown itself was concerning.
Such unknowns include how information is used ($n = \num{13}$), how much information is collected ($n = \num{8}$), and who has access to the information ($n = \num{6}$). 
The response of P123 (\emph{somewhat concerned}) illustrates this: \enquote{I am somewhat concerned considering that I do not really know how much information Google collects from me, or if I can limit how much in any meaningful way.}

Privacy fatigue was another theme described by \num{18}~participants who stated they were resigned to the loss of privacy when using online websites and services such as Google.
For example, P55 (\emph{not at all concerned}) noted: \enquote{Not much I can do to stop it, so why worry?}
Participant P88 (\emph{slightly concerned}) added:
\enquote{I believe that data gathering and analytics is occurring in all aspects of life from the television that you watch to the groceries that you buy so I'm really not that concerned and there is little I can do to block the collecting of information.}

Another \num{15} participants considered Google's data collection as a reasonable trade-off for free applications and services.
For instance, P62 (\emph{not at all concerned}) noted: \enquote{They aren't collecting any more from me than from anyone else and it is the cost of using their free services.} 
Participant P60 (\emph{slightly concerned}) articulated that the conveniences of Google services made them willing to accept the data collection, explaining:
\enquote{Google provides many conveniences to me as an internet user. It's not ideal that so much information is gathered about me, but I'm not prepared to give up these conveniences.}

There were \num{19} participants whose concerns were primarily for the security of their collected data.
Examples include security concerns about their data being released in some way ($n = \num{12}$), such as a data breach at Google.
Like when P9 (\emph{moderately concerned}) explained, \enquote{I'm concerned that my data may fall into the wrong hands through a breach or other methods.}
Also, \num{11} participants were concerned that their data could be mishandled or misused in a way that would be harmful.
An instance of this is when P72 (\emph{somewhat concerned}) mentioned, \enquote{It exposes my information to those out there that could end up being malicious.}

However, \num{29} participants reported being unconcerned about the amount of information Google collects.
Eight participants stated that the information Google collected was non-sensitive in nature, like P67 (\emph{not at all concerned}), who recalled, \enquote{I don't have anything that I would deem important that I don't mind Google knowing.}
Another five participants felt the collection of information was low risk to them personally, \eg, P33, who reported being \emph{not at all concerned}, said simply, \enquote{As long as it isn't harmful to me personally, I am okay with it.}
Still, four participants claimed they had nothing to hide.
For instance, P107 (\emph{somewhat concerned}) explained, \enquote{I need it, and I don't have anything to hide, so it doesn't really matter to me.} 
Two participants were unconcerned because they had never experienced negative impacts, \eg, P105 (\emph{slightly concerned}) mentioned, \enquote{Only slightly concerned because I've never had any bad experiences with Google or privacy.}
\num{16} participants explicitly noted that they trusted Google for the reason they were unconcerned.

\paragraph{Initial Perceptions of Benefits}
Participants seemed uncertain when first asked how often they benefit (\ref{app:survey:q3}) from the amount of information Google collects (\cf \autoref{fig:q3-q10-q12-bar}; \emph{top bar}).
Almost half stated they \emph{sometimes} benefit from data collection, and the remaining  tend to  choose \emph{rarely} or \emph{never} ($n = \num{56}$) instead of \emph{often} or \emph{always} ($n = \num{32}$).
This result indicates that direct benefits coming from the data collection are intangible for the participants.

When explaining why or why not they benefit (\ref{app:survey:q3a}), \num{57} participants described the benefits of personalized advertisements.
Better recommendations were emphasized by \num{27} participants, \eg, when using \gsearch ($n = \num{16}$), \yt ($n = \num{8}$), or Maps ($n = \num{7}$).
Also, \num{16} participants mentioned an overall improved experience.
For instance, P139, who reported benefiting \emph{sometimes} from the amount of information collected, responded:
\enquote{I believe that the info collected is used to improve my experience with Google products. For example, using my watch history for better recommendations on YouTube.}
Participant P66 (\emph{sometimes}), who found personalized advertisements useful, said, 
\enquote{I see ads pop up that are relevant to me quite frequently. I'm assuming Google know what ads to show me based on information they've collected on me.}
Nine participants found it useful to revisit their Google activity history, such as when P16 (\emph{sometimes}) had this to say,
\enquote{It will remember your history to be able for easy access next time you want to visit the same page.}
\num{10} participants mentioned a personalized experience, \eg, P50 (\emph{sometimes}), who said, \enquote{It can be used to personalize apps and websites, or to track things like my location, which can be useful to me.}
Five participants called out the benefit of Google~Chrome filling out forms automatically and remembering passwords for websites.
As an example of this, P48 (\emph{sometimes}) explained, \enquote{When I am autofilling forms with information Google has already collected from me I benefit by saving time.}

However, \num{24} participants stated that personalized ads were not beneficial to them.
This was highlighted by P172 (\emph{rarely}), who shared:
\enquote{I don't find the personalized ads useful at all. Instead, I find them creepy because it means Google is watching my every step online.}

\paragraph{Changing Perceptions of Concerns}
\begin{table}[t]
\centering
\caption[Wilcoxon tests]{\label{tab:concern-benefit-tests}
Two-tailed Wilcoxon signed-rank tests of level of concern (\ref{app:survey:q2}, \ref{app:survey:q9}, \ref{app:survey:q11}) and frequency of benefit (\ref{app:survey:q3}, \ref{app:survey:q10}, \ref{app:survey:q12}).
The tests show a significant \textbf{decrease} in the level of concern after participants saw the interest from their \adsettings page.
For the frequency of benefit, the tests yield only an \textbf{increase} of benefit after the labeling task but only for a 10\% significance level.
}
\footnotesize
\begin{tabular*}{\columnwidth}{
@{}
l
l
@{\extracolsep{\fill}}
r
r
@{\extracolsep{0pt}~}
l
@{\extracolsep{\fill}}
r
r
@{\extracolsep{0pt}~}
l
@{}
}
\toprule
\multicolumn{2}{@{}l@{}}{\textbf{Concern}} &
\multicolumn{1}{@{}c@{}}{\textbf{n}} &
\multicolumn{2}{@{}c@{}}{\textbf{Effect Size}} &
\multicolumn{1}{@{}c@{}}{\textbf{W}} &
\multicolumn{2}{@{}c@{}}{\textbf{Pr(\textgreater{}\textbar{}W\textbar{})}} \\
\midrule
 \ref{app:survey:q2} &  \ref{app:survey:q9} & 170 & 0.009 & (small) & 848.0 & 0.668 & \\
 \ref{app:survey:q2} & \ref{app:survey:q11} & 133 & \textbf{0.405} & \textbf{(mod.)} & \textbf{1995.0} & \textbf{\textless{}0.001} & *** \\
 \ref{app:survey:q9} & \ref{app:survey:q11} & 133 & \textbf{0.388} & \textbf{(mod.)} & \textbf{1373.5} & \textbf{\textless{}0.001} & *** \\
\midrule
\multicolumn{2}{@{}l@{}}{\textbf{Benefit}} &
\multicolumn{1}{@{}c@{}}{\textbf{n}} &
\multicolumn{2}{@{}c@{}}{\textbf{Effect Size}} &
\multicolumn{1}{@{}c@{}}{\textbf{W}} &
\multicolumn{2}{@{}c@{}}{\textbf{Pr(\textgreater{}\textbar{}W\textbar{})}} \\
\midrule
 \ref{app:survey:q3} & \ref{app:survey:q10} & 170 & 0.157 & (small) & 854.5 & 0.069 & $\cdot$ \\
 \ref{app:survey:q3} & \ref{app:survey:q12} & 133 & 0.098 & (small) & 557.0 & 0.454 & \\
\ref{app:survey:q10} & \ref{app:survey:q12} & 133 & 0.097 & (small) &  85.5 & 0.186 & \\
\bottomrule
\end{tabular*}
\footnotesize
\textbf{Signif. codes:} $\text{***}~\widehat{=} < 0.001$; $\text{**}~\widehat{=} <0.01$; $\text{*}~\widehat{=} <0.05$; $\cdot~\widehat{=} <0.1$
\end{table}

 \begin{figure*}[thbp]
  \begin{subfigure}[t]{\columnwidth}
    \includegraphics[width=\textwidth]{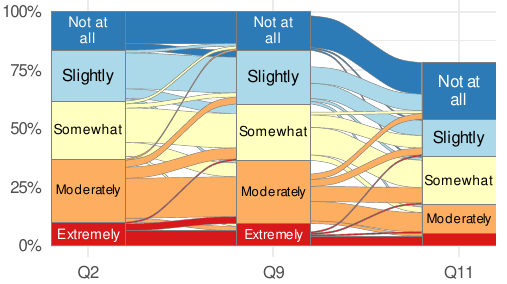}
    \caption[Perceived benefit alluvium plot]{\protect\label{fig:q2-q9-q11-alluvium}
      Level of concern alluvium plot.
    }
  \end{subfigure}
  \hfill
  \begin{subfigure}[t]{\columnwidth}
    \includegraphics[width=\textwidth]{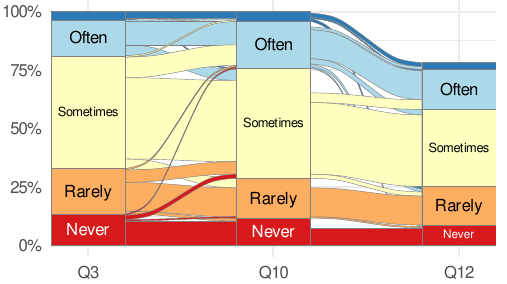}
    \caption[Frequency of benefit alluvium plot]{\protect\label{fig:q3-q10-q12-alluvium}
      Frequency of benefit alluvium plot.
    }
  \end{subfigure}
  \caption[Alluvium plots of level of concern and frequency of benefit]{\protect\label{fig:concern-benefit-alluvium}
    Changes in the participants' assessment of
    (a) the level of concern (\ref{app:survey:q2}, \ref{app:survey:q9}, \& \ref{app:survey:q11}) and 
    (b) the frequency of benefit (\ref{app:survey:q3}, \ref{app:survey:q10},  \& \ref{app:survey:q12})
    at the beginning of the survey, after the labeling tasks, and after exposure to the inferred interests.
  }
\end{figure*}

Comparing only the proportions in \autoref{fig:q2-q9-q11-bar} (\emph{top two bars}), it seems that there is no change between the initial  concern (\ref{app:survey:q2}) and concern after the labeling tasks (\ref{app:survey:q9}). This is confirmed by a two-tailed Wilcoxon signed-rank test (\cf \autoref{tab:concern-benefit-tests}, comparing \ref{app:survey:q2} to \ref{app:survey:q9});  no significant difference was found ($W=844.0;p=0.668$). 
More details on how the participants change their level of concern are shown in the alluvium plot in \autoref{fig:q2-q9-q11-alluvium} (\emph{first two columns}).
Out of the 170 participants, \num{114} did not change their initial assessment after doing the tasks.
The remaining 56 participants changed their assessment: \num{27} moved to a lower concern level, and \num{29} moved up.

Compared to our previous work~\cite{farke-21-privacy-dashboads}, this is notable as back then, we found a decrease in the level of concern after participants observed their Google My~Activity dashboard.
Here, we observe no significant difference before and after observing activities and applying ad~interests to those activities. This could be due to the further reasoning requested of participants, where they now had to consider what Google may actually learn from their activities.
However, after reviewing their true interest as assigned by Google (\ie, a \enquote{summarized} version of their activities), their concern actually dropped.
We discuss these further results in more detail in \autoref{sec:results:reactions}.

\begin{table}[t]
\centering
\caption[Post-exposure concern regression analysis]{\label{tab:q2-q9-regression}
Generalized logistic regression model to describe the likelihood of a \textbf{decrease} in the level of concern before (\ref{app:survey:q2}) and after the labeling tasks (\ref{app:survey:q9}).
The Aldrich-Nelson pseudo $R^2$ of the model is \num{0.2883}.
}
\footnotesize
\begin{tabular*}{\columnwidth}{
@{}
l
@{\extracolsep{\fill}}
rrr
@{\extracolsep{0pt}~}
l
@{}
}
    \toprule
    \textbf{Factor} &
    \textbf{Est.} &
    \multicolumn{1}{@{}c@{}}{\textbf{OR}} &
    \multicolumn{2}{@{}c@{}}{\textbf{Pr(\textgreater{}\textbar{}z\textbar{})}} \\
    \midrule
    (Intercept)                                                  & -19.01 & 5.55E-09 & 0.986 &     \\
    $\text{\ref{app:survey:q2}} = \textit{Extremely concerned}$  & 18.75  & 1.39E+08 & 0.986 &     \\
    $\text{\ref{app:survey:q2}} = \textit{Moderately concerned}$ & 18.52  & 1.10E+08 & 0.987 &     \\
    $\text{\ref{app:survey:q2}} = \textit{Somewhat concerned}$   & 17.05  & 2.54E+07 & 0.988 &     \\
    $\text{\ref{app:survey:q2}} = \textit{Slightly concerned}$   & 15.57  & 5.79E+06 & 0.989 &     \\
    Labeling difference                                & 0.26   & 1.30E+00 & 0.869 &     \\
    Increasing benefit                                           & \textbf{2.13}  & \textbf{8.39E+00} & \textbf{\textless{}0.001} & *** \\
    $\text{Gender} = Male$                                       & \textbf{-1.47}  & \textbf{2.31E-01} & \textbf{0.016}               & *   \\
    $\text{Age} \in\{18-34,~25-34\}$                             & -0.56  & 5.73E-01 & 0.445 &     \\
    $\text{Age} \in\{35-44,~45-54\}$                             & 0.80   & 2.22E+00 & 0.269 &     \\
    $\text{Edu.} \in \{\textit{No sch.g, (Sm.) HS}\}$            & -0.27  & 7.66E-01 & 0.734 &     \\
    $\text{Edu.} \in \{\textit{Sm. col., Assoc., Prof.}\}$       & -0.31  & 7.37E-01 & 0.596 &     \\
    $\text{Has S.T.E.M major}$                                   & 0.18   & 1.20E+00 & 0.778 &     \\
    \bottomrule
\end{tabular*}
\footnotesize
\textbf{Signif. codes:} $\text{***}~\widehat{=} < 0.001$; $\text{**}~\widehat{=} <0.01$; $\text{*}~\widehat{=} <0.05$; $\cdot~\widehat{=} <0.1$
\end{table}

While there was no significant difference between concerns, overall, there were participants who did change their concern level, both up and down.
We conducted a generalized logistic regression to identify the factors that more likely lead to a decrease in concern.
Logistic regression is an ideal model for this analysis as it considers a single binary outcome: \emph{After the labeling task, did the participant decrease their concern, or not?} (\ref{app:survey:q2} vs.\ \ref{app:survey:q9}).
The model provides how correlated the input factors are to a binary variable, as well as how much more likely a given input would lead to the output factor (the odds ratio).
For input factors, we selected 1) demographic input factors, such as age, education, and gender, as these were used in prior work~\cite{farke-21-privacy-dashboads}; 2) how differing the participants' labeling of activity were compared to how Google applied labels; 3) and finally, what their initial concern levels were as represented in their Likert response to \ref{app:survey:q2}, where \emph{Not at all Concerned} is withheld and included in the intercept.
Note, the same description applies to the results in \autoref{tab:q3-q10-regression}, but in this case, regarding benefit instead of concern.

Regarding a decrease in the level of concern (\autoref{tab:q2-q9-regression}), we find that participants who increase their benefit rating (comparing \ref{app:survey:q3} to \ref{app:survey:q10}) were $8.39\times$ more likely to decrease their concern assessment ($\beta=2.13$; $OR=8.39$; $p<0.001$).
Participants who identified as male were more likely to increase their concern, $4.33\times$ higher ($\beta=-1.47$; $OR=0.231$; $p=0.016$).

Participants were also asked (\ref{app:survey:q9a}) why they changed or did not change their assessment of how concerned they are with the amount of information Google collects about their activities online.
Participants who changed their perspective and became more concerned ($n = \num{33}$) after completing the labeling task included those who were worried about the amount of information collected ($n = \num{18}$) or that their search history was recorded ($n = \num{7}$), or that the data collected could be used to track them online ($n = \num{6}$). 
For instance, P26, who reported being \emph{somewhat concerned} before the labeling task and changed to \emph{moderately concerned} afterward, responded, \enquote{I forgot how many little details of the websites I was visiting were being collected.}
P64 (\emph{somewhat concerned} to \emph{moderately concerned}) explained, \enquote{I didn't realize what a snapshot of my life that they keep. And these are just recent searches.}
Participant P93 (\emph{slightly concerned} to \emph{somewhat concerned}), reacting to an activity collected in the 18-month or older timeframe, illustrated:
\enquote{It's kind of surprising that they can essentially take screenshots of everything that I look up and store it. Also, I was kind of surprised to see search results from 2020, which makes me think about how much information they could be storing and from how long ago.}

Learning that Google collected search history made participant P53 (\emph{somewhat concerned} to \emph{moderately concerned})  more concerned as they noted, \enquote{I didn't know Google collected my entire search history.}
Participant P83 (\emph{somewhat concerned} to \emph{moderately concerned}) felt the data collection was invasive and that they were being tracked as they explained, \enquote{It feels invasive knowing how much I am being tracked at all times as if I have no privacy and can be tracked and found by anyone at anytime.}
Others increased their concern when they found that the information collected was sensitive in nature ($n = \num{2}$), very specific to them ($n = \num{2}$), and contained location data ($n = \num{2}$). 
For example, P79, who was \emph{slightly concerned} and became \emph{moderately concerned}, highlighted, \enquote{It doesn't seem to track as much variety of information as I thought and a lot of what I saw it tracked was very personal.}
P28 (\emph{not at all concerned} to \emph{slightly concerned}) added, \enquote{I noticed that specific addresses were collected that I am not comfortable sharing.} 

In contrast, some participants who changed their perception of the data collection became less concerned ($n = \num{9}$) after completing the labeling task.
Their reasons for becoming less concerned included that the information shown was not sensitive ($n = \num{9}$) or was generic in nature ($n = \num{1}$). 
For instance, participant P5, who was \emph{somewhat concerned}  and changed to \emph{not at all concerned} said simply, \enquote{There is nothing sensitive in this information.} 
While P99 (\emph{slightly concerned} to \emph{not at all concerned}) added:
\enquote{I realized how little of importance the results that came up where, most of them were in regards to what I was currently studying or just watching on YouTube which had no real value to a person outside of recommending me things that would be more to my preferences.}

\paragraph{Changing Perceptions of Benefits}
The top two bars in \autoref{fig:q3-q10-q12-bar} suggest a slight  increase in the frequency of initial benefit assessment (\ref{app:survey:q3}) and after the labeling tasks (\ref{app:survey:q10}). A detailed overview of the pair-wise changes of benefit can be found in  \autoref{fig:q3-q10-q12-alluvium} (\emph{left \& middle column}).
From the initial benefit (\ref{app:survey:q3}) to the benefit after the labeling tasks (\ref{app:survey:q10}), \num{32} participants increased their ratings.
A majority of \num{120} participants did not change, and \num{18} participants even decreased their assessments.

We performed a two-tailed Wilcoxon signed-rank test ($W=854.5;p=0.069$), which is trending toward significant with a possible small effect. This is not a significant result, \ie, $p<0.05$, but the test was two-tailed.
A one-tailed test would be significant; however, this was outside our analysis plan. We had no priors on which directions we should expect a change.
More so, our test is only sufficiently powered to identify moderate effects ($w=0.3$) with $n=170$, and the effect size is most likely small.
A much larger sample, \eg, with over \num{1500}~participants, would be needed to properly test for small effects ($w=0.1$), but this sample suggests that such a result could be the case in a larger evaluation.

\begin{table}[t]
\centering
\caption[Post-exposure frequency of benefit regression analysis]{\label{tab:q3-q10-regression}
Generalized logistic regression model to describe the likelihood of an \textbf{increase} in the frequency of benefit before (\ref{app:survey:q3}) and after the labeling tasks (\ref{app:survey:q10}).
The Aldrich-Nelson pseudo $R^2$ of the model is \num{0.1696}.
}
\footnotesize
\newsavebox{\equbox}
\begin{tabular*}{\columnwidth}{
@{}
l
@{\extracolsep{\fill}}
rrr
@{\extracolsep{0pt}~}
l
@{}
}
    \toprule
    {\textbf{Factor}} &
    {\textbf{Est.}} &
    \multicolumn{1}{@{}c@{}}{\textbf{OR}} &
    \multicolumn{2}{@{}c@{}}{\textbf{Pr(\textgreater{}\textbar{}z\textbar{})}} \\
    \midrule
    (Intercept)                                            & -16.70 & 5.60E-08 & 0.986 &   \\
    $\text{\ref{app:survey:q3}} = \textit{Never}$          & 15.22  & 4.06E+06 & 0.988 &   \\
    $\text{\ref{app:survey:q3}} = \textit{Rarely}$         & 15.63  & 6.11E+06 & 0.987 &   \\
    $\text{\ref{app:survey:q3}} = \textit{Sometimes}$      & 15.17  & 3.89E+06 & 0.988 &   \\
    $\text{\ref{app:survey:q3}} = \textit{Often}$          & 13.35  & 6.28E+05 & 0.989 &   \\
    Labeling difference                          & -0.19  & 8.29E-01 & 0.884 &   \\
    Decreasing concern                                     & \textbf{1.11}   & \textbf{3.02E+00} & \textbf{0.026} & * \\
    $\text{Gender} = Male$                                 & 0.52   & 1.68E+00 & 0.247 &   \\
    $\text{Age} \in\{18-34,~25-34\}$                       & -0.51  & 6.02E-01 & 0.388 &   \\
    $\text{Age} \in\{35-44,~45-54\}$                       & -0.76  & 4.65E-01 & 0.203 &   \\
    $\text{Edu.} \in \{\textit{No sch.g, (Sm.) HS}\}$      & 0.29   & 1.34E+00 & 0.661 &   \\
    $\text{Edu.} \in \{\textit{Sm. col., Assoc., Prof.}\}$ & -0.11  & 9.00E-01 & 0.824 &   \\
    $\text{Has S.T.E.M major}$                             & 0.51   & 1.66E+00 & 0.307 &   \\
    \bottomrule
\end{tabular*}
\footnotesize
\textbf{Signif. codes:} $\text{***}~\widehat{=} < 0.001$; $\text{**}~\widehat{=} <0.01$; $\text{*}~\widehat{=} <0.05$; $\cdot~\widehat{=} <0.1$
\end{table}

 To complement the regression analysis of the level of concern, as described above, we calculated a similar logistic regression with an outcome variable for an increase in the benefit scale. 
The model in \autoref{tab:q3-q10-regression} shows symmetry with the prior regression as the only significant factor is \enquote{decreasing concern.}
That is, participants who decrease their level of concern are roughly $3.02\times$ more likely to increase their perceived benefit rating, which mirrors the prior result where participants with an increase in benefit were more likely to decrease their concern.

These results complement the findings of our previous study~\cite{farke-21-privacy-dashboads}, where we found significant increases in benefit paired with decreases in concern after participants viewed their own Google activities.
Here, there is still a tight relationship between decreases in concern and increases in benefit, but overall shifts are weaker.

After completing the labeling task, participants were again asked (\ref{app:survey:q10a}) to explain why they changed or did not change their assessment of how often they benefit from the amount of information Google collects about them.
Many participants ($n = \num{28}$) changed their perspective and increased their perceived benefit from data collection following the labeling task. 
Among these, common themes included improved search results ($n = \num{6}$), product personalization ($n = \num{5}$), better recommendations ($n = \num{5}$), and more accurate targeted ads ($n = \num{3}$). 
Participant P63, who changed from \emph{sometimes} to \emph{often} benefiting from Google's data collection, said: \enquote{Google collects information and uses some of it to aid in my searches, narrow down searches, and provide information relevant to me.}
P20 (\emph{sometimes} to \emph{often}) added, \enquote{I believe that Google understands what kind of person I am and what I want.}

Conversely, those five participants who became aware of fewer benefits of data collection after the labeling task cited targeted ads not being beneficial ($n = \num{3}$) and a lack of improvement in the experience of using Google search and products ($n = \num{2}$).  
For instance, participant P174, who went from \emph{rarely} down to \emph{never} benefiting from Google's data collection, replied, \enquote{I do not believe personalized ads are worth trading my personal information.}
Moreover, participant P101 (\emph{sometimes} to \emph{rarely}) explained, \enquote{I changed my assessment here because I cannot recall any instance of the data shown being used to improve my internet user experience.}

\subsection*{\emph{Summary of RQ2 [Impact on Benefit/Concern]}}
We find a near-even split between participants who are concerned about Google's data collection and those who are not.
Participants described to benefit of personalized advertisements and better recommendations.
Contrary to the results of our previous work~\cite{farke-21-privacy-dashboads}, we have not observed any significant change in the level of concern or benefit before and after the labeling tasks, suggesting that connecting activities with implied interests better maintains prior perceptions.

\subsection{\ref{RQ:reactions-interests}: Reactions to Ad~Interests}\label{sec:results:reactions}

We asked the 133~participants with interests on their \adsettings page additional questions regarding their concerns that Google applied these interests (\ref{app:survey:q11}), the perceived benefit of \enquote{personalization} (\ref{app:survey:q12}), and how accurate the labels are in representing them as an individual (\ref{app:survey:q13}).
The 37~participants without \adsettings pages skipped this part of the survey.

\paragraph{Concerns of \adsettings Interests}
The results for the level of concern~(\ref{app:survey:q11}) are found in \autoref{fig:q2-q9-q11-bar} and show a shift toward \emph{not at all}.
More than half of the participants are either \emph{slightly} ($n=27$) or \emph{not at all} concerned ($n=41$) after viewing their Google \adsettings page.
About a quarter of the participants ($n=35$) are still \emph{somewhat} concerned.
Another \num{21} participants are \emph{moderately} concerned, and only a minority of \num{9} participants are \emph{extremely} concerned.
\autoref{fig:q2-q9-q11-alluvium} shows that from question \ref{app:survey:q9} to \ref{app:survey:q11}, both on concern, fewer participants ($n = \num{75}$) did not change their assessment.
\num{37}~participants had disabled Google's ad~interests.
Of the remaining participants, \num{46} decreased their rating, and \num{12} increased it.

Again, we performed two-tailed Wilcoxon signed-rank tests to determine if there were significant differences before and after viewing their Google-applied labels. These results are found in \autoref{tab:concern-benefit-tests}. While there was no significant change after the labeling tasks, viewing the true labels led to significant \emph{decreases} in concern both from \ref{app:survey:q2} ($W=1995$; $p<0.001$) and \ref{app:survey:q9} ($W=1373.5$; $p<0.001$) compared to \ref{app:survey:q11}. When viewing the true labels, which they viewed as more general and, at times, inaccurate (discussed below), their concern about how the advertising interests were assigned significantly dropped.

Participants provided short-answer responses about their change in concern (\ref{app:survey:q11a}). Many ($n = \num{58}$) responses included concerns about personal privacy regarding data collected by Google.
Among these participants, concerns about the amount of information collected were common ($n = \num{16}$). 
For instance, participant P24, who was \emph{moderately concerned} about Google learning their interest highlighted:
\enquote{They know a lot about me. I like the personalization aspects that come with them having that data, but I'm not sure if I'm completely ok with them knowing all that about me.}
Participant P84 (\emph{extremely concerned}) added: \enquote{The information shown was a lot of items collected about me.}

Concerns about Google selling data were again common ($n = \num{9}$), such as P62, who reported being \emph{slightly concerned} and said, \enquote{I am concerned that they may sell the data to companies that I don't want to have it.}
Furthermore, P119 (\emph{moderately concerned}) felt that the Google user community is not being fairly compensated for the use of their data:
\enquote{Again, it is the concept that Google and other giants tech companies don't pay the community for the right to use their data. Make these companies pay data rents to the community on top of guaranteeing autonomy.}
Seven participants were concerned about third-party access to their interests, such as when P19 said, \enquote{I am concerned that my interests could be shared elsewhere/with other parties, which may not be desirable.}

Another concern, mention by six participants, was the number of interests that Google learns, e.g., P93 (\emph{slightly concerned}) reported:
\enquote{I am just slightly concerned because of how many interests they have me tagged in. Kind of makes them know more than a friend that you talk to everyday would.}

Six participants were concerned about how much Google knows about them.
P57 (\emph{extremely concerned}) shared, \enquote{Yikes, like I said, they know everything about me.} 
Additionally, there were participants ($n = \num{4}$) who felt concerned about being tracked, such as P36 (\emph{slightly concerned}), who illustrated, \enquote{I just don't like having everything about me tracked. It's really bizarre to someone who is 55 years old.}

Security concerns ($n = \num{6}$) were also described by participants who were worried about their data being misused ($n = \num{3}$) or released ($n = \num{3}$).
For instance, participant P59 (\emph{slightly concerned}) shared, \enquote{It's a little concerning just to have that much information concentrated in one place/on one platform, because you never know how it could be used or who could potentially access it.}

On the other hand, many ($n = \num{52}$) participants were unconcerned after viewing their Google-inferred ad~interests and felt they were generic in nature ($n = \num{10}$), and the fact that Google inferred this knowledge was low risk ($n = \num{12}$).
Like participant P68 (\emph{slightly concerned}), who described:
\enquote{If anything, I thought all the interests shown were much more generic than I previously thought. It doesn't really reveal anything I'd consider personal.}
Participant P145 (\emph{not at all concerned}) replied, \enquote{I don't care if they know my interests, I can't see any harm coming from that.}

Many participants ($n = \num{21}$) found Google's inferred interests to be inaccurate, leading to less concern.
Participant P74, who was \emph{not at all concerned}, illustrated:
\enquote{Because a lot is so wrong it doesn't matter. It's an annoyance to have \enquote{targeted} ads which are way off.}

\paragraph{Perceived Benefit of \adsettings Interests}
In the final frequency of benefit question (\ref{app:survey:q12}), a smaller proportion of the participants ($n=\num{56}$) as compared to the first two questions \ref{app:survey:q3}, \ref{app:survey:q10} stated to \enquote{sometimes} benefit from the interests Google is inferring from their activity (\cf \autoref{fig:q3-q10-q12-bar}).
One third of the participants thought that they \enquote{rarely} ($n=\num{28}$) or even \enquote{never} ($n=\num{15}$) benefit.
The remaining participants stated to benefit either \enquote{often} ($n=\num{29}$) or \enquote{always} ($n=\num{5}$) from Google's data collection.
As shown in \autoref{fig:q3-q10-q12-alluvium}, \num{75} of the participants did not change their benefit assessment from \ref{app:survey:q10} to \ref{app:survey:q12}.
Of the remaining people, a majority decreased ($n=\num{46}$), and only \num{12} participants increased their rating.
Also, as mentioned before, \num{37} did not answer this question. And, as shown in \autoref{tab:concern-benefit-tests}, there was no significant change for \ref{app:survey:q12}.
Likely, participants already solidified their view on the benefits of the advertising interests, and viewing their true labels only confirmed their priors.

In explaining their change (or lack thereof) of benefit (\ref{app:survey:q12a}), many ($n = \num{88}$) described benefits around the idea that Google learns their interests. 
Among these participants, a common concept ($n = \num{27}$) was the better recommendations that they received when using Google products and services. 
According to participant P76, who \emph{sometimes} benefits, \enquote{Knowing more about me can influence the recommendations I'm given which impacts my user experience positively.}
Another common idea ($n = \num{27}$) was a more personalized experience when using Google products and services, such as when P164 (\emph{sometimes}) said, \enquote{Sometimes relevant products, ads, and websites are presented, and that's beneficial to me and helps make my internet experience more personalized.}
Some ($n = \num{27}$) participant responses included the benefits of targeted advertising. For instance, P68 (\emph{sometimes}) noted, \enquote{I think the interests were mostly correct, so I like having ads tailored to me.}
Others ($n = \num{15}$) mentioned they benefit from improved Google search results. 
For example, participant P63 (\emph{often}) explained, \enquote{Google's learning helps them to cater searches and information to my needs, makes everything more relevant.}
Seven participants noted improvements in their \yt experience, such as participant P14 (\emph{often}), who replied, \enquote{Google learning about my interests helps me to get YouTube videos and websites recommended to me based on my interests so that I know I'll always like what is recommended to me.}

Conversely, \num{33}~participants reported little or no benefits from Google learning their interests.
Common themes included a dislike for personalized ads ($n = \num{7}$), and Google learning interests was unhelpful ($n = \num{5}$). 
For instance, P17 (\emph{never}) said, \enquote{I do not benefit from google learning my interests, I do not need them to help me find things to buy.}
And P72 (\emph{rarely}) added, \enquote{I don't see this as being beneficial in terms of what it can offer.} 

\begin{figure}[t]
  \centering
  \includegraphics[width=\columnwidth]{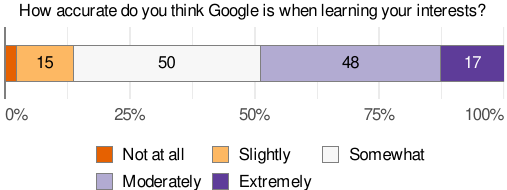}
  \caption[Perceived Accuracy after \adsettings]{\label{fig:q13-bar}
    Perceived accuracy after seeing the list of interests from the \adsettings page.
  }
\end{figure}

\paragraph{Perceived Accuracy of \adsettings Interests}
Participants were asked (\ref{app:survey:q13} and \ref{app:survey:q13a}) to assess and explain how accurate they viewed Google's assigned interests.
The results of \ref{app:survey:q13}, presented in \autoref{fig:q13-bar}, suggest that most participants consider their \adsettings interests to be accurate.
Of the \num{133} participants who had \adsettings enabled, \num{65} stated that the interests were \enquote{moderately} or \enquote{extremely} accurate.
Another \num{50}~participants were more ambivalent about the accuracy with a \enquote{somewhat} rating, and \num{15}~participants stated that the interests are \enquote{slightly} accurate.
Only a minority of three participants found the interest applied to their accounts to be \enquote{not at all} accurate.

Still, \num{54} participants explained that Google had some incorrect inferences. 
Among these participants, some complained about the interests being based on one-time-activities or searches, such as when P29, who rated the interests as \emph{somewhat accurate}, described:
\enquote{Some of my searches are specific while others are just one time events and sometime a search I do for someone else. Google can't tell the difference and therefore never leverages any info collected in any way meaningful to me.}
Other examples were given by \num{13} participants, such as P51 (\emph{somewhat accurate}), who recalled, \enquote{There were things listed, like Anime and Bread Makers, that I've never looked at before.}
\num{12} participants reported that their \adsettings{} page had many incorrect inferences, such as P157 (\emph{not at all accurate}), who complained about out-of-date interests, \enquote{These collections are all old, I have moved on.}

In contrast, \num{22} participants reported that many of their interests were accurate, and \num{24} participants even said most were accurate.
For instance, P57, who found the interests to be \emph{extremely accurate}, exclaimed, \enquote{Because I use them for everything, they know more about me than I know about myself, and that's not a joke.}
Participant P101 (\emph{moderately accurate}) was able to \emph{connect the dots} from Google's inferred interests back to past online activities when they explained:
\enquote{The interests that Google identified as my interests were very accurate and I am able to assume that they have concluded these as my interests directly through my interactions with the internet.}
This theme was also reflected in participant P85's response, \enquote{I think Google picks up on things I search a lot and places I shop and dine at so Google knows me better than I thought.}
Another \num{20} participants found only some of the interests to be correct, like P24 (\emph{moderately accurate}), who shared, \enquote{There are a few things that are off, but overall, it's a lot more accurate than I expected.}
Three participants even found the interests to accurately reflect their personality, e.g., P32 (\emph{extremely accurate}) said, \enquote{I saw so many corners of my personality and identity in the listed categories.}

\subsection*{\emph{Summary of RQ3 [Reactions to Ad Interests]}}
Overall, participants found the interests applied by Google to be accurate, with some exceptions.
In contrast to the labeling task, we find that viewing the labels that Google applied significantly decreased participants' concern, mainly because participants perceived the labels to be too generic or that they were low-risk labels or inaccurate.

\section{Discussion}\label{sec:discussion}
In this paper, we evaluated the impact of \enquote{connecting the dots} of the data food chain by asking participants to supply their own advertising labels to their own real online activities. Through our study, we got the following answers to our research questions:
\begin{enumerate}[leftmargin=*,label=RQ\arabic*,format=\bfseries,topsep=5pt,itemsep=5pt,partopsep=0pt,parsep=0pt,labelwidth=2pt]
    \item \emph{[Understanding of Ad Interests]} Participants applied interests to their own activities differently than Google, mostly being either too generic or overly specific.
    \item \emph{[Impact on Benefit/Concern]} We found no significant shifts in concern or benefit before and after the labeling task.
    \item \emph{[Reactions to Ad Interests]} Viewing their true ad~interest, as assigned by Google, causes a significant decrease in participants' concerns. The interests are accurate but perceived as generic and low risk, contributing to the decrease of concern.
\end{enumerate}

The strongest insight gained is the potential to build next-gen\-er\-ation privacy dashboards that both provide users the ability to actively manage their data collection and also access to \emph{how} that data may be used, such is in the behavioral advertising ecosystem. Such a system has the potential to better maintain priors of concern and benefit expectations, as opposed to existing dashboards, which just display raw information, that has been shown to reduce concern and increase perceived benefit~\cite{farke-21-privacy-dashboads}. After the labeling task, participants expressed what labels they believe Google would apply given their activities, their priors were reinforced:
\begin{enumerate*}[label=\alph*)]
    \item If they were concerned about data collection, they remained concerned, or
    \item if they were unconcerned, they remained unconcerned.
\end{enumerate*}

However, there is a potentially large shortcoming when moving to such a new privacy dashboard design due to the mismatched expectations in the specificity of the advertising labels that users imagine, compared to the rather generic labels used by advertisers. We refer to this phenomenon as the \emph{generic paradox}, and the impact was that after participants viewed their actual labels on Google's \adsettings page, the results reverted back to be in line with our previous study~\cite{farke-21-privacy-dashboads}, we observed a large decrease in concern and a moderate increase in benefits. It is critical that new interfaces both connect the dots of data collection and provide important context to counteract the generic paradox. One design direction is to build interfaces that help users moderate their expectations and better explain how the collected data will be used to generate actionable insights and profiles, as well as to demonstrate how even seemingly low-risk generic labels can actually be combined to unique identifiers that can lead to abuse.

In the rest of this section, we explore this theme and other insights gained from this research and how they generally apply to the current online behavioral advertising ecosystem.

\paragraph{Methodological Shift Compared to Prior Work}
Our prior work~\cite{farke-21-privacy-dashboads} performed a pre-post study design with an intervention that had people interact directly with the activities that Google collected about them via the Google My~Activity dashboard.
At the time, we found both a significant decrease in participants' concern and an increase in benefit from Google's data collection.
The study in this work has a similar pre-post study design.
However, the intervention goes to a higher order of data processing and asks participants to reason, by completing a labeling task, about how Google uses their collected activities to infer information about them via ad~interests.
This methodological shift of the intervention also led to a shift in concern and benefit findings.
The results show a stable pre-post concern rating and a minor increase in benefit.
The more modest increase in benefit could be the result of the intervention where the focus is on the benefits of behavioral advertising (seen as less beneficial) than when considering activity history, which reminds individuals of the quality of Google services, such as YouTube and \gsearch.

According to our previous study~\cite{farke-21-privacy-dashboads}, the significant reduction in concern after viewing the My~Activity page was partly due to the sense of control over the collected data.
In contrast, at the time we conducted the study described in this paper, Google's \adsettings page offered close to no opportunities to customize, manage, and understand inferred ad~interests.
The decrease in concern we observed with participants viewing their interests could be attributed to a lack of perceived risk posed by the interests.
For example, many participants indicated that even though the labels were accurate, they were generic and less specific than they feared.
Google has since revised the dashboard---now called \emph{My~Ad~Center} (\href{https://myadcenter.google.com}{myadcenter.google.com})~\cite{google-22-my-ad-center}---in an effort to make it more useful for managing interests.

Still, there is an unresolved question of the best strategy for assisting people in managing the data collected about them and how that data is used.
Providing additional controls via data dashboards eases concerns and helps to demonstrate the benefit of data collection, but it still may not fully explain why such controls should be used or raise awareness about the associated risks.
In some ways, these results suggest that a better explanation of data flows can help expose people and reinforce existing concerns, but future work is needed to fully understand how to properly present such data collection mechanisms and processing in a data dashboard.

\paragraph{The Generic Paradox}
According to our qualitative findings, participants were commonly unconcerned or became less concerned about Google's data collection because their inferred ad~interests were generic.
For instance, the ad~interest \enquote{Apparel} could apply to anyone with clothing-related activities so this interest could be applied to many people.
However, given the fact that participants had a median of 180 inferred interests and that almost half of them reported that the interests were \emph{extremely} or \emph{moderately} accurate, it is important to consider what happens when numerous discrete generic (but accurate) pieces of information are combined to form a more complete and precise picture of an individual.
Paradoxically, by focusing on the generic quality of individual interests and not on the volume and accuracy of those interests, people will be less likely to manage or remove their interests and will be more likely to allow Google to continue collecting data about their online activities.
A recent study by Reitinger et al.~\cite{reitinger-23-ad-settings} found that while many of their participants shared general topics (``Shopping''), a lot of interests were unique to each participant and highly individualized.
Furthermore, it only takes 33~bits of entropy to identify an individual uniquely among the world's population~\cite{narayanan-08-robust}, and de-anonymization attack techniques have been successful in targeted advertisements~\cite{korolova-10-microtargeted-ads, venkatadri-18-data-broker}, and in many other domains~\cite{zang-11-location-data, montjoye-15-credit-card, caliskan-18-binaries}. 

\paragraph{People's Understanding of Inferencing}
Prior work on ad inferencing suggests that people base their understanding on direct, one-to-one connections between past online behaviors and inferred interests~\cite{rader-20-inferences-vs-perceptions}. Few understand the abilities of machine learning to aggregate data from millions of individuals to create detailed personal inferences~\cite{warshaw-16-intuitions}.
For instance, if a person searches for \enquote{pants,} they simply connect that with the ad~interest \enquote{Apparel,} while few take the extra step to match \enquote{lawnmower} with \enquote{Homeownership Status: Homeowners.}
Our findings show that people have misconceptions when assigning ad~interests based on their past online activities, often because they are too literal and overly specific, such as naming specific brands and exact locations.
The actual Google ad~interests, on the other hand, tend to focus on grouping peoples' interests into broad categories, with the goal to appeal to a wide range of advertisers thereby increasing Google's revenue.
Qualitative data suggests that participants understood that interests are inferred from their interactions with Google products, but little understanding exists about how those interactions are combined with other data.

\paragraph{The Data Food Chain}
The metaphor of a data food chain~\cite{nissenbaum-19-data-food-chain} helps to expose the information hierarchy in which data of a higher order (\ie, ad~interests) is a function of data of a lower order (\ie, activities).
For instance, the inferred ad~interest \emph{Sports Cars} might be a function of the \yt activity where a person watched a video of a \enquote{\emph{2022 Mazda CX-30,}} combined with other lower-order data such as location, past searches, and the inferred interests of other people on Google with similar activity profiles.
This data aggregation raises the question of whether consent to data collection can be meaningful, informed, and explicit when privacy disclosure mechanisms such as privacy policies provide transparency primarily regarding lower-order data collection but limited transparency regarding what, how, and why higher-order data is derived.
Compounded by the fact that people do not fully understand what is inferred about them, as is evidenced by our findings that participants self-assigning interests differently, this questions the usefulness of privacy notices in the context of higher-order data processing.
In a report filed in the U.S. Department of Justice anti-trust complaint against Google~\cite{doj-22-google-antitrust-litigation}, privacy expert Martin argues that \enquote{certain uses of Google search profile data for the purpose of returning personalized advertising would constitute a \enquote*{non-contextual, secondary} data use that would violate users' privacy expectations.}

\paragraph{A Lack of Alternatives}
When participants were asked which other products (than Gmail) they would use (\ref{app:survey:s3}), we observed Google's products and services dominating. 
In our biased sample of \num{170} Google account owners we considered in our analysis, Chrome was used by 168, \yt by 167, Gmail by 164, \gmaps by 151, and \gsearch by 148, emphasizing Google's profound impact on the Web.
Since Google's primary source of revenue is advertising, and the current business model of targeted advertising is based on collecting an ever-increasing amount of behavioral data, it is incumbent that people spend more and more time engaged with their products and services.
With all these services, a \emph{free} Google account adds enormous value.
P207 said, \enquote{Google provides many conveniences to me as an Internet user.}
These conveniences, unfortunately, force people to make a trade-off between their privacy and the products they have come to rely on and make competition hard to come by.
As P131 summarized, \enquote{I still value some Google products over their competition so I compromise on the collection of my activity information.}

\paragraph{My~Ad~Center Design Changes}
Google revised \adsettings in November 2022 to provide more \enquote{transparency and control}~\cite{google-22-my-ad-center}.
In comparison to \adsettings, the new dashboard lists fewer interests, now called \enquote{topics} (\enquote{Jewelry}), and introduces a new tab for managing \enquote{brands} (\enquote{Amazon}), as well as demographic and aggregated interests, now called \enquote{categories} (\enquote{Homeownership Status}).
With the new dashboard, users have better control over the number of ads that are based on a certain topic by liking or disliking the topic.
Unfortunately, this design change can only make ads more relevant but does not improve users' privacy.
An interesting design change is the new \enquote{recent ads} section, where Google itself has started connecting advertisements with interests.
Here, for every ad they have seen, users can check which topic an advertiser used for segmenting and targeting them.
Exploring if this new design impacts perceived benefits or concerns, as well as if it encourages users to take more actions in managing their privacy online, would be a fruitful area of future research.

\clearpage
\makeatletter
\interlinepenalty=10000
\bibliographystyle{ACM-Reference-Format}
\bibliography{main.bib}
\makeatother

\nobalance
\clearpage

\appendix
\section{Appendix}\label{sec:appendix}

\subsection{Survey Instrument}\label{app:survey}

\setlength{\columnsep}{0pt}
\setlength{\multicolsep}{0pt}
\setlength{\parindent}{0pt}
\definecolor{structure}{HTML}{03588C} %
\definecolor{note}{HTML}{038080} %

\begin{scriptsize}
Thank you for your interest in our survey. Your answers are important to us. \textbf{Please read the following instructions carefully:}

\begin{enumerate*}[label=(\roman*)]
    \item Take your time in reading and answering the questions.
    \item Answer the questions as accurately as possible.
    \item It is okay to say that you don't know an answer.
\end{enumerate*}

\begin{center}
\textcolor{note}{\emph{[A horizontal rule, like below, indicates a new page in the questionnaire.]}}

\hrulefill
\end{center}

\begin{questions}[label=S\arabic*]
    \item Do you have a \textbf{personal Gmail address} (an email address ending in \enquote{gmail.com})?
    \begin{multicols}{2}
        \begin{answers}
            \item Yes
            \item No
        \end{answers}
    \end{multicols}
    \label{app:survey:s1}

    \item \textbf{How long} do you have that Gmail address?
    \begin{multicols}{2}
        \begin{answers}
            \item Less than a year
            \item One year
            \item Three years
            \item Five years
            \item More than five years
            \item I do not have a Gmail address
            \item Unsure
        \end{answers}
    \end{multicols}
    \label{app:survey:s2}
    
    \item\label{app:survey:s3} Which \textbf{other Google products} do you currently use? (Select all that apply.)
    \begin{multicols}{3}
        \begin{answers}
            \item Gmail
            \item Google Maps
            \item YouTube
            \item Google Chrome
            \item Google Search
            \item Google Play
            \item Google Drive
            \item Google News
            \item Google Pay
            \item Android device
            \item None of these
        \end{answers}
    \end{multicols}
    
\end{questions}
\hrulefill

\begin{questions}[label=S\arabic*, resume]
    \item \textbf{How frequently} do you \textbf{use} these \textbf{products}?
    \label{app:survey:s4}
    \textcolor{note}{\emph{[Included only products selected in \ref{app:survey:s3}. If \enquote{None of these} was selected question was hidden.]}}

    \smallskip
    
    \begin{tabular*}{\linewidth}{@{}l@{\extracolsep{\fill}}*{6}{c@{}}}
        \toprule
                       & Always     & Often      & Sometimes  & Rarely     & Never      & Unsure     \\
        \midrule
        Gmail          & $\bigcirc$ & $\bigcirc$ & $\bigcirc$ & $\bigcirc$ & $\bigcirc$ & $\bigcirc$ \\
        Google Maps    & $\bigcirc$ & $\bigcirc$ & $\bigcirc$ & $\bigcirc$ & $\bigcirc$ & $\bigcirc$ \\
        YoutTube       & $\bigcirc$ & $\bigcirc$ & $\bigcirc$ & $\bigcirc$ & $\bigcirc$ & $\bigcirc$ \\
        Google Chrome  & $\bigcirc$ & $\bigcirc$ & $\bigcirc$ & $\bigcirc$ & $\bigcirc$ & $\bigcirc$ \\
        Google Search  & $\bigcirc$ & $\bigcirc$ & $\bigcirc$ & $\bigcirc$ & $\bigcirc$ & $\bigcirc$ \\
        Google Play    & $\bigcirc$ & $\bigcirc$ & $\bigcirc$ & $\bigcirc$ & $\bigcirc$ & $\bigcirc$ \\
        Google Drive   & $\bigcirc$ & $\bigcirc$ & $\bigcirc$ & $\bigcirc$ & $\bigcirc$ & $\bigcirc$ \\
        Google News    & $\bigcirc$ & $\bigcirc$ & $\bigcirc$ & $\bigcirc$ & $\bigcirc$ & $\bigcirc$ \\
        Google Pay     & $\bigcirc$ & $\bigcirc$ & $\bigcirc$ & $\bigcirc$ & $\bigcirc$ & $\bigcirc$ \\
        Android device & $\bigcirc$ & $\bigcirc$ & $\bigcirc$ & $\bigcirc$ & $\bigcirc$ & $\bigcirc$ \\
        \bottomrule
    \end{tabular*}

    \item \textbf{How important} is using \textbf{Google products} to your Internet experience?
    \begin{multicols}{2}
        \begin{answers}
            \item Not important
            \item Slightly important
            \item Moderately important
            \item Important
            \item Very important
        \end{answers}
    \end{multicols}
    \label{app:survey:s5}
\end{questions}
\hrulefill
\subsubsection*{Please Install the \enquote{Survey Assistant} Extension}
On this page, we will ask you to install our browser extension, which will assist you during the study. The installation takes only a few steps. Please follow the instructions below. We will assist you to uninstall the browser extension at the end of the survey.

\hrulefill
\subsubsection*{Google Sign-in}
This survey also requires that you login to \textbf{your primary Google account} for accessing items in your My Activity and Ad Settings page. My Activity and Ad Settings are services provided by Google to make you see more useful ads and offer a more personalized experience, including faster searches and automatic recommendations. Please use an account with an email address that ends in \textbf{@gmail.com}.

\smallskip
\textbf{Privacy Note:}
We do not track or store your email address as part of this study, and we will not be able to tie your email address to any results or analysis. The researchers will never see your email address. At no time do the researchers have access to your Google account.

\hrulefill

\begin{questions}
    \item\label{app:survey:q1} \textbf{How aware} are you of the amount of information that Google is collecting about your activities online?
    \begin{multicols}{2}
        \begin{answers}
            \item Not at all aware
            \item Slightly aware
            \item Somewhat aware
            \item Moderately aware
            \item Extremely aware
        \end{answers}
    \end{multicols}
    \item\label{app:survey:q2} \textbf{How concerned} are you with the amount of information Google is collecting about your activities online?
    \begin{multicols}{2}
        \begin{answers}
            \item Not at all concerned
            \item Slightly concerned
            \item Somewhat concerned
            \item Moderately concerned
            \item Extremely concerned
        \end{answers}
    \end{multicols}
    \begin{questions}
        \item\label{app:survey:q2a} Please explain why.
        \begin{answers}[widest=Answer:]
            \item[Answer:] \rule{3cm}{.4pt}
        \end{answers}
    \end{questions}
    
    \item\label{app:survey:q3} \textbf{How often} do you \textbf{benefit} from the amount of information that Google collects about your activities online?
    \begin{multicols}{2}
        \begin{answers}
            \item Never
            \item Rarely
            \item Sometimes
            \item Often
            \item Always
        \end{answers}
    \end{multicols}
    
    \begin{questions}
        \item\label{app:survey:q3a} Please explain why.
        \begin{answers}[widest=Answer:]
            \item[Answer:] \rule{3cm}{.4pt}
        \end{answers}
    \end{questions}
\end{questions}

\hrulefill

\subsubsection*{What is Google \adsettings}
The following video briefly introduces you to Google's \adsettings page. For every account, Google provides a settings page called \adsettings, which allows you to control whether and how personalized ads are shown to you. \textbf{Please watch this 1-minute video before proceeding:}
\begin{center}
    \textcolor{note}{\emph{[YouTube video describing Google's \adsettings.]}}
    \href{https://www.youtube.com/watch?v=hdBTv1x7iLk}{\includegraphics[width=0.95\columnwidth]{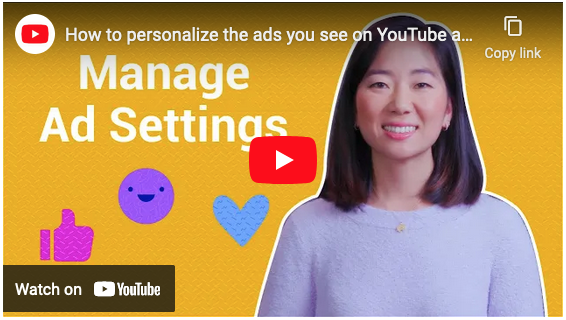}}
\end{center}

\hrulefill

\subsubsection*{Prepare Survey}
Please click the \enquote{Prepare Survey} button to proceed to the next page.

\textcolor{note}{\emph{[Interactive checklist, which \emph{visualized} the progress of gathering data from the My~Activity and Ad~Settings page.]}}

\bigskip
\begin{itemize}[nosep]
    \item Check for browser extension
    \item Check for ad interests (Number of found interests)
    \item Check for Google Search activities (Number of found Google~Search activities)
    \item Check for YouTube activities (Number of found YouTube activities)
    \item Check for Google Maps activities (Number of found Google~Maps)
\end{itemize}

\hrulefill

\subsubsection*{Task Instructions}
Next, we will ask you to assign at least 3 interests you think Google will learn from the shown activity, which we randomly selected from your Google My~Activity page. This task is repeated with up to \textbf{9} different activities.

\smallskip
\textbf{Note:}
We will store your assigned interests and activities. Please make sure to only assign interests to activities you feel comfortable sharing with us. You can skip any activity by clicking \enquote{Skip this Activity.}

Before we start, let us explain how you can assign these interests on the next page.

\subsubsection*{Task Demo}
\begin{center}
\textcolor{note}{\emph{[Interactive tutorial of the Labeling Task user interface.]}}
\end{center}
\begin{enumerate}[nosep,noitemsep]
    \item \textbf{The Activity:} On the next page, you will be presented with one of your activities picked from your Google My~Activity page.
    \item \textbf{Skipping an Activity:} If you feel uncomfortable with the shown activity you can skip the activity by pressing this button.
    \item \textbf{Reason for Skipping:} If you choose to skip, please select why you do not like to assign an interest and share the activity with us.
    \item \textbf{Assign an Interest:} To assign an interest you think Google will learn from the shown activity above, use this text box to enter your idea.
    \item \textbf{List of Interests:} While you type, we will display some suggestions that Google typically applies. You can select one by clicking on it.
    \item \textbf{Selected Interests:} Your selected interests will be displayed in the text box and can be removed by pressing on the \enquote{X} icon.
    \item \textbf{Custom Interests:} If no suggestion fits your idea, you can also enter your own interests.
    \item \textbf{Add Custom Interest:} To add a custom interest, please press the plus button or the \enquote{Enter} key on your keyboard.
    \item \textbf{Finished:} To start your assignment task, please click the Next button to proceed to the next page.
\end{enumerate}

\subsubsection*{Labeling Tasks}
In the next part, we will ask you questions about \textbf{nine activities} from your My~Activity page. The activities are chosen randomly. We do not collect information about that activity as part of this survey. That information remains private, \textbf{only accessible to you} and Google. We only note which service the activity is associated with, \eg, \enquote{Google search} vs. \enquote{YouTube view,} and the date on which it occurred. Further details are not collected as part of this survey.

\bigskip
\begin{center}
   \fbox{\includegraphics[width=0.8\columnwidth]{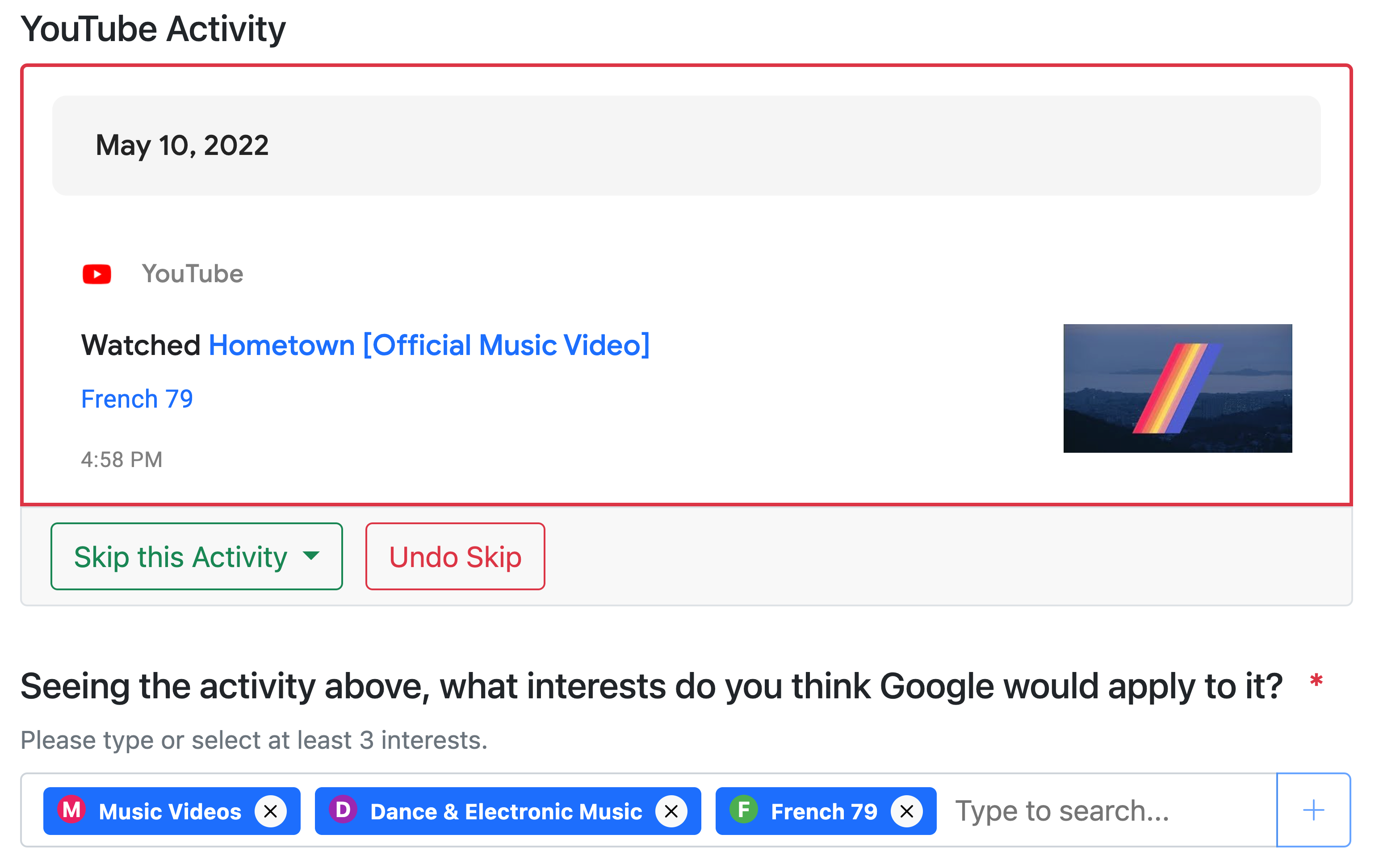}}
\end{center}
    
\begin{questions}

    \item[\textbf{I}]\label{app:inference} \textbf{Seeing the activity above, what interests do you think Google would apply to it?}
    \begin{center}
    \textcolor{note}{\emph{[This question was repeated up to 9 times for different services and time periods.]}}
    \end{center}
\end{questions}

\hrulefill

Now that you have worked with activities and interests, we would like to repeat some of the questions from the beginning of the survey.
\begin{questions}[start=9]
    \item\label{app:survey:q9} \textbf{How concerned} are you with the amount of information Google is collecting about your activities online?
    
    \begin{multicols}{2}
        \begin{answers}
            \item Not at all concerned
            \item Slightly concerned
            \item Somewhat concerned
            \item Moderately concerned
            \item Extremely concerned
        \end{answers}
    \end{multicols}
    \begin{questions}
        \item\label{app:survey:q9a} Please explain why you changed or not changed your assessment.
        \begin{answers}[widest=Answer:]
            \item[Answer:] \rule{3cm}{.4pt}
        \end{answers}
        
    \end{questions}

    \item\label{app:survey:q10} \textbf{How often} do you \textbf{benefit} from the amount of information that Google collects about your activities online?
    \begin{multicols}{2}
        \begin{answers}
            \item Never
            \item Rarely
            \item Sometimes
            \item Often
            \item Always
        \end{answers}
    \end{multicols}
    
    \begin{questions}
        \item\label{app:survey:q10a} Please explain why you changed or not changed your assessment.
        \begin{answers}[widest=Answer:]
            \item[Answer:] \rule{3cm}{.4pt}
        \end{answers}
    \end{questions}
\end{questions}

\subsubsection*{Google \adsettings}
Next, we will show you the interests from your Google \adsettings page. These interests are assigned to your account by Google.

\smallskip
\textbf{Note:}
Please take your time to look through the interests. You may proceed to the next page after 30 seconds.

\hrulefill

\subsubsection*{Presentation of Interests}

\begin{center}
    \textcolor{note}{\emph{[Participants were presented with the interest from their \adsettings page.]}}

    \fbox{\includegraphics[width=0.8\columnwidth]{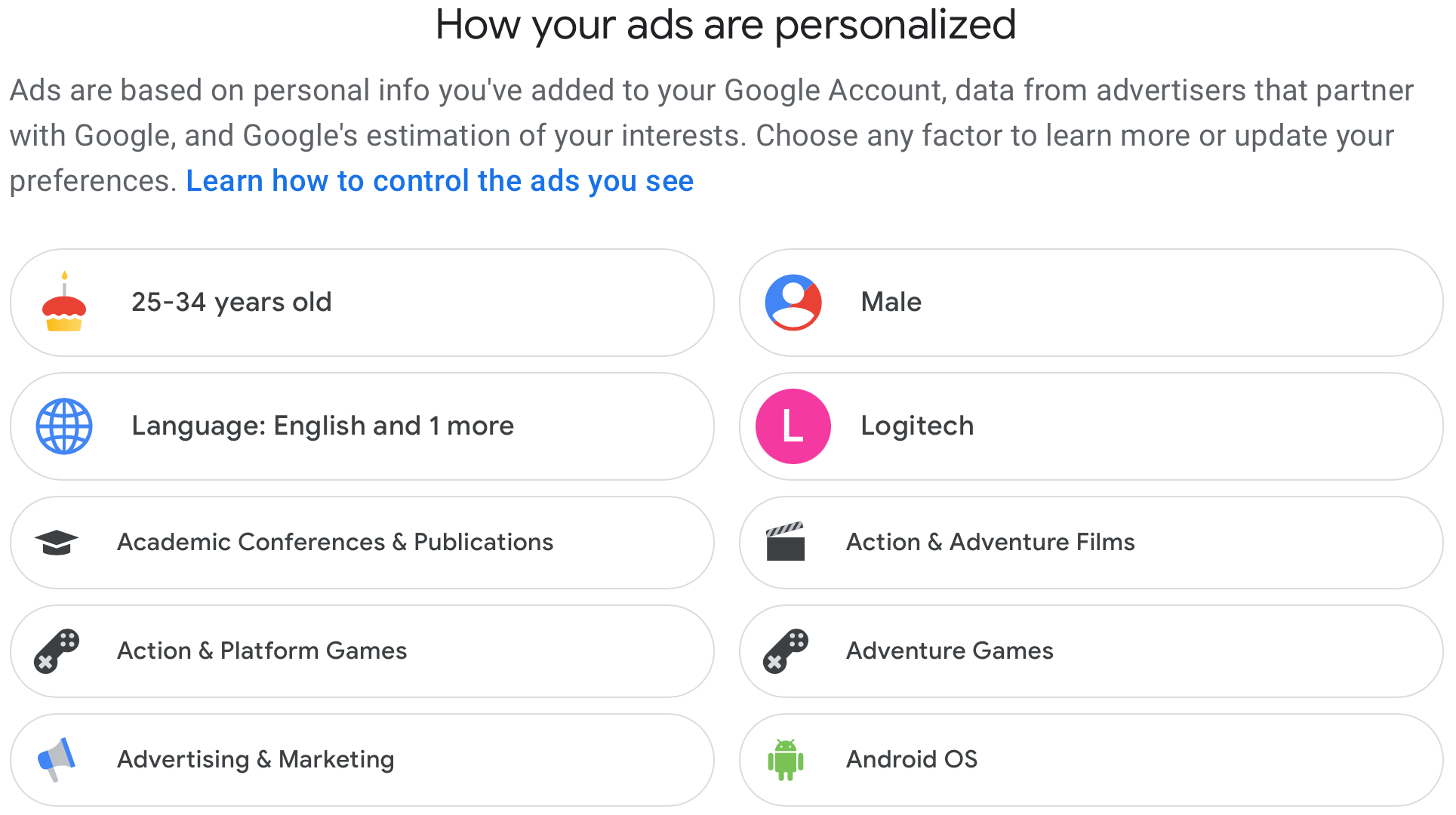}}
\end{center}

\hrulefill

\begin{questions}[resume]
    \item\label{app:survey:q11} \textbf{How concerned} are you about Google learning about your interests based on your activities online?
    \begin{multicols}{2}
        \begin{answers}
            \item Not at all concerned
            \item Slightly concerned
            \item Somewhat concerned
            \item Moderately concerned
            \item Extremely concerned
        \end{answers}
    \end{multicols}
    \begin{questions}
        \item\label{app:survey:q11a} Please explain why.
        \begin{answers}[widest=Answer:]
            \item[Answer:] \rule{3cm}{.4pt}
        \end{answers}
    \end{questions}
    
    \item\label{app:survey:q12} \textbf{How often} do you \textbf{benefit} from Google learning your interests based on your activities online?
    \begin{multicols}{2}
        \begin{answers}
            \item Never
            \item Rarely
            \item Sometimes
            \item Often
            \item Always
        \end{answers}
    \end{multicols}
    
    \begin{questions}
        \item\label{app:survey:q12a} Please explain why.
        \begin{answers}[widest=Answer:]
            \item[Answer:] \rule{3cm}{.4pt}
        \end{answers}
    \end{questions}

    \item\label{app:survey:q13} \textbf{How accurate} do you think Google is when learning your interests?
    \begin{multicols}{2}
        \begin{answers}
            \item Not at all accurate
            \item Slightly accurate
            \item Somewhat accurate
            \item Moderately accurate
            \item Extremely accurate
        \end{answers}
    \end{multicols}
    
    \begin{questions}
        \item\label{app:survey:q13a} Please explain why.
        \begin{answers}[widest=Answer:]
            \item[Answer:] \rule{3cm}{.4pt}
        \end{answers}
    \end{questions}
\end{questions}

\newpage
\hrulefill

\begin{questions}[label=D\arabic*]
    \item\label{app:survey:d1} What is your \textbf{gender}?
    \begin{multicols}{2}
        \begin{answers}
            \item Woman
            \item Man
            \item Non-binary
            \item Prefer not to disclose
            \item Prefer to self-describe
        \end{answers}
    \end{multicols}
    
    \item\label{app:survey:d2} What is your \textbf{age}?
    \begin{multicols}{2}
        \begin{answers}
            \item 18 -- 24
            \item 25 -- 34
            \item 35 -- 44
            \item 45 -- 54
            \item 55 -- 64
            \item 65 or older
            \item Prefer not to disclose
        \end{answers}
    \end{multicols}
    
    \item\label{app:survey:d3} Are you of Hispanic, Latino, or Spanish \textbf{origin}?
    \begin{answers}
        \item No, not of Hispanic, Latino, or Spanish origin
        \item Yes, Mexican, Mexican Am., Chicano
        \item Yes, Puerto Rican
        \item Yes, Cuban
        \item Yes, another Hispanic, Latino, or Spanish origin
        \begin{itemize}[nosep,noitemsep]
            \item Type, for example, Salvadoran, Dominican, Colombian, Guatemalan, Spaniard, Ecuadorian, etc.: \rule{3cm}{.4pt}
        \end{itemize}
        
    \end{answers}
    
    \item\label{app:survey:d4} What is your \textbf{race}?
    \begin{multicols}{2}
    \begin{answers}
        \item White
        \item Black or African American
        \item American Indian or Alaskan Native~~
        \item Chinese
        \item Filipino
        \item Asian Indian
        \item Vietnamese
        \item Korean
        \item Japanese
        \item Other Asian
        \item Native Hawaiian
        \item Samoan
        \item Chamorro
        \item Other Pacific Islander
    \end{answers}
    \end{multicols}

    \item\label{app:survey:d5} Which of these describes your personal \textbf{income} last year?
    \begin{multicols}{2}
    \begin{answers}
        \item No income
        \item \$1 to \$9,999
        \item \$10,000 to \$24,999
        \item \$25,000 to \$49,999
        \item \$50,000 to \$74,999
        \item \$75,000 to \$99,999
        \item \$100,000 to \$149,999
        \item \$150,000 and greater
        \item Prefer not to disclose
    \end{answers}
    \end{multicols}
    
    \item\label{app:survey:d6} What is the highest degree or \textbf{level of school} you have completed?
    
    \begin{answers}
        \item No schooling completed
        \item Some high school, no diploma
        \item High school graduate, diploma, or equivalent
        \item Some college credit, no degree
        \item Trade/technical/vocational training
        \item Associate degree
        \item Bachelor's degree
        \item Master's degree
        \item Professional degree (\eg, J.D., M.D.)
        \item Doctorate degree
        \item Prefer not to disclose
        \item Other (please specify)
    \end{answers}

    \item\label{app:survey:d7} Do you have a \textbf{major} in one of the following fields?
    \begin{multicols}{2}
    \begin{answers}
        \item S.T.E.M.
        \item Humanities
        \item Law
        \item Social science
        \item Prefer not to disclose
        \item Other (please specify)
    \end{answers}
    \end{multicols}

    \item\label{app:survey:d8} What is your \textbf{marital status}?
    
    \begin{multicols}{2}
        \begin{answers}
            \item Married
            \item Living as married
            \item Divorced
            \item Widowed
            \item Separated
            \item Single, never been married
            \item Prefer not to disclose
            \item Prefer to self-describe
        \end{answers}
    \end{multicols}

    \item\label{app:survey:d9} How many \textbf{children} do you have?
    
    \begin{multicols}{2}
        \begin{answers}
            \item 0
            \item 1
            \item 2--4
            \item 4+
            \item Prefer not to disclose
        \end{answers}
    \end{multicols}

    \item\label{app:survey:d10} Which of the following best represents \textbf{how you think of yourself}?
    \begin{answers}
        \item Lesbian or gay
        \item Straight, that is, not lesbian or gay
        \item Bisexual
        \item Something else
        \item I don't know the answer
        \item Prefer not to disclose
        \item Prefer to self-describe
    \end{answers}
\end{questions}
\end{scriptsize}
\clearpage

\onecolumn
\subsection{Tables}\label{app:tables}

\begin{table}[!h]
\centering
\caption[Participant Demographics]{\label{tab:demographics}
Demographic data of the participants.
We collected this data at the end of survey (\ref{app:survey:d1}--\ref{app:survey:d10}: Optional Questions).
}
\footnotesize
\renewcommand{\arraystretch}{0.6}
\begin{tabular*}{\columnwidth}{
>{\bfseries}l
l
@{}
@{\extracolsep{\fill}}
r
r
r
r
}
\toprule
& &
\multicolumn{2}{c}{\textbf{Completed}} &
\multicolumn{2}{c}{\textbf{Eligible}} \\
& &
\multicolumn{2}{c}{\emph{(n = \num{235})}} &
\multicolumn{2}{c}{\emph{(n = \num{170})}} \\
\cmidrule{3-6}
&
&
\multicolumn{1}{r}{\textbf{n}} &
\multicolumn{1}{r}{\textbf{\%}} &
\multicolumn{1}{r}{\textbf{n}} &
\multicolumn{1}{r}{\textbf{\%}} \\
\midrule
Gender
& Man                     & 130 & 55 & 98 & 58 \\
& Woman                   &  94 & 40 & 67 & 39 \\
& Non-binary              &   6 &  3 &  4 &  2 \\
& Preferred not to answer &   5 &  2 &  1 &  1 \\
\midrule
Age
& 18--24                  & 21 &  9 & 16 &  9 \\
& 25--34                  & 77 & 33 & 61 & 36 \\
& 35--44                  & 60 & 25 & 46 & 27 \\
& 45--54                  & 32 & 14 & 22 & 13 \\
& 55--64                  & 30 & 13 & 19 & 11 \\
& 65 or older             &  9 &  4 &  5 &  3 \\
& Preferred not to answer &  6 &  2 &  1 &  1 \\
\midrule
Education
& Bachelor's degree                        & 91 & 39 & 61 & 36 \\
& Some college credit, no degree           & 45 & 19 & 39 & 23 \\
& High school diploma, or equivalent       & 28 & 12 & 21 & 12 \\
& Associate degree                         & 26 & 11 & 21 & 12 \\
& Master's degree                          & 19 &  8 & 12 &  7 \\
& Trade / technical / vocational training  &  7 &  3 &  4 &  2 \\
& Doctorate degree                         &  6 &  2 &  5 &  3 \\
& Some high school, no diploma             &  4 &  2 &  3 &  2 \\
& Professional degree (\eg, J.D., M.D.)    &  3 &  1 &  2 &  1 \\
& Preferred not to answer                  &  6 &  3 &  2 &  1 \\
\midrule
Income
&                                                    No income  &  6 &  3 &  4 &  2 \\
&      \SI{1}[\dollar]{\relax} to   \SI{9999}[\dollar]{\relax}  & 30 & 13 & 24 & 14 \\
&  \SI{10000}[\dollar]{\relax} to  \SI{24999}[\dollar]{\relax}  & 35 & 15 & 27 & 16 \\
&  \SI{25000}[\dollar]{\relax} to  \SI{49999}[\dollar]{\relax}  & 59 & 25 & 43 & 25 \\
&  \SI{50000}[\dollar]{\relax} to  \SI{74999}[\dollar]{\relax}  & 38 & 16 & 24 & 14 \\
&  \SI{75000}[\dollar]{\relax} to  \SI{99999}[\dollar]{\relax}  & 27 & 11 & 22 & 13 \\
& \SI{100000}[\dollar]{\relax} to \SI{149999}[\dollar]{\relax}  & 19 &  8 & 14 &  8 \\
& \SI{150000}[\dollar]{\relax} and greater                      &  8 &  3 &  4 &  2 \\
&                                       Preferred not to answer & 13 &  6 &  8 &  5 \\
\bottomrule
\end{tabular*}
\end{table}

\begin{table}[!h]
\caption[Top interest labels]{
Top 10 interests collected from the participants.
}
\label{tab:top-interests}

\footnotesize
\centering
\begin{tabular*}{\columnwidth}{@{}r@{\extracolsep{\fill}}l@{}l@{}r@{}r@{}}
\toprule
\textbf{\#} &
\textbf{Topic} &
\textbf{Interest} &
\multicolumn{1}{@{}c@{}}{\textbf{n}} &
\multicolumn{1}{@{}c@{}}{\textbf{\%}} \\
\midrule
 1 & Shopping                 & Shopping                          & 125 & 94 \\
 2 & Business \& Industrial   & Business Services                 & 123 & 92 \\
 2 & Games                    & Computer \& Video Games           & 123 & 92 \\
 3 & News                     & News                              & 122 & 92 \\
 3 & Travel \& Transportation & Travel \& Transportation          & 122 & 92 \\
 4 & Computers \& Electronics & Computers \& Electronics          & 120 & 90 \\
 4 & Online Communities       & Social Networks                   & 120 & 90 \\
 5 & Books \& Literature      & Books \& Literature               & 119 & 89 \\
 5 & Food \& Drink            & Cooking \& Recipes                & 119 & 89 \\
 6 & Arts \& Entertainment    & Comics \& Animation               & 118 & 89 \\
 6 & Food \& Drink            & Restaurants                       & 118 & 89 \\
 7 & Internet \& Telecom      & Mobile Phones                     & 117 & 88 \\
 8 & Sports                   & Sports                            & 115 & 86 \\
 9 & Shopping                 & Apparel                           & 113 & 85 \\
 9 & Shopping                 & Flowers                           & 113 & 85 \\
10 & Arts \& Entertainment    & Celebrities \& Entertainment News & 112 & 84 \\
\bottomrule
\end{tabular*}
\end{table}

 \begin{small}
\begin{table}[!h]
  \centering
  \footnotesize
  \caption{\label{tab:example-interests}
    Example interests (as shown on the \adsettings page).
  }
    \begin{tabular*}{\textwidth}{l@{\extracolsep{\fill}}llll}
    \toprule
    \multicolumn{1}{l}{\textbf{Demographic}} & \multicolumn{1}{@{\extracolsep{\fill}}l}{\textbf{Advertiser}} & \multicolumn{1}{l}{\textbf{Aggregated}} & \multicolumn{1}{l}{\textbf{Activity}} & \multicolumn{1}{l}{\textbf{YTC}} \\
    \midrule
    25-34 years old    & Best Buy    & Company Size:           Small Employer (1-249 E.) & American Football     & A video from Chromebook \\
    Female               & Capital One & Education Status:       Bachelor's Degree         & Camera Lenses         & A video from Coinbase   \\
    Language: English  & Etsy        & Homeownership Status:   Homeowners                & Cooking \& Recipes    & A video from HBO Max    \\
                       & Facebook    & Household Income:       Lower Middle              & Guitars               & A video from Lyft       \\
                       & Target      & Job Industry:           Technology Industry       & Movie \& TV Streaming & A video from Pizza Hut  \\
                       & Walmart     & Marital Status:         In a Relationship         & Porsche               & A video from Target     \\
                       & Wayfair     & Parental Status:        Not A Parent              & Yoga \& Pilates       & A video from TurboTax   \\
    \bottomrule
    \end{tabular*}%
\end{table}%

 \end{small}

\subsection{Codebook}\label{app:codebook}

\begin{small}
\begin{itemize}[nosep]

\item\textbf{collection-beneficial~(319)}

\emph{personalized-ads~(106), better-recommendations~(85), personalization~(57), search~(46), improved-experience~(34), youtube~(22), revisit-activities~(14), maps~(10), autofill-forms~(3), search-history~(2), news~(2), storing-information~(2), free-products-services~(2), store-passwords~(2), google-rewards~(1), sometimes~(1), saved-logins~(1), websites~(1)}

\item\textbf{privacy-concerns~(218)}

\emph{amount-of-information~(62), selling-data~(33), information-collection~(27), tracking~(24), sensitive-information~(19), third-parties~(18), monitoring~(14), do-not-want-others-to-see~(12), invasion-of-privacy~(11), information-used-against-me~(9), amount-of-interests~(6), amount-of-knowledge~(6), location~(4), search-history~(3), scary~(3), targeted-advertising~(3), creepy~(3), ads-too-specific~(2), specific-information~(2), societal-consequences~(1), stalking~(1), what-information-collected~(1), interests-too-accurate~(1), behavior-modification~(1), ads-everywhere~(1), uncomfortable-sharing~(1), how-long-data-stored~(1), why-information-collected~(1)}

\item\textbf{unconcerned~(125)}

\emph{non-sensitive~(31), low-risk~(21), information-is-generic~(11), interests~(7), nothing-bad-has-happened~(5), collection-meets-expectations~(4), data-collection~(4), comfortable-sharing~(4), because-interests-not-accurate~(3), interests-not-sensitive~(3), has-read-privacy-policy~(2), do-not-care-who-knows-interests~(1), information-is-secured-by-google~(1), information-collected-is-not-important~(1)}

\item\textbf{no-benefit~(116)}

\emph{no-use-ads~(32), personalized-ads-not-helpful~(6), interests-not-helpful~(5), information-collected-not-helpful~(3), no-use-recommendations~(2), experience-not-enhanced~(2), causes-problems~(1), recommendations-not-helpful~(1), not-paid-for-data-collection~(1), personalized-ads-invasive~(1)}

\item\textbf{did-not-change-perspective~(109)}

\item\textbf{some-interests-incorrect~(54)}

\emph{one-off-activities~(15), gave-examples~(13), outdated~(3), does-not-like-wrong-interests~(2), is-okay~(1), conflicting-interests~(1), very-inaccurate~(1), wants-to-delete~(1)}

\item\textbf{unknowns~(38)}

\emph{how-information-used~(17), who-has-access~(10), what-information-collected~(8), how-much-information~(4), why-information-collected~(2), how-long-information-stored~(1), where-information-goes~(1), risk~(1)}

\item\textbf{interests-not-accurate~(32)}

\emph{only-one-search~(3), wrong-context~(1)}

\item\textbf{security-concerns~(30)}

\emph{data-misuse~(14), data-released~(10), data-breach~(9), identity-theft~(3)}

\item\textbf{privacy-tradeoff~(27)}

\emph{free-services~(19), benefits~(1), necessary~(1)}

\item\textbf{trust-google~(26)}

\emph{transparent~(1)}

\item\textbf{privacy-resigned~(25)}

\item\textbf{most-interests-correct~(24)}

\emph{very-accurate~(1), better-than-expected~(1), based-on-activities~(1)}

\item\textbf{privacy-aware~(23)}

\item\textbf{many-interests-correct~(22)}

\emph{based-on-activities~(4), assumptions~(1)}

\item\textbf{some-interests-correct~(20)}

\emph{but-vague~(1), lack-of-understanding~(1)}

\item\textbf{many-interests-incorrect~(12)}

\emph{gave-examples~(3), outdated~(2), one-off-activities~(1), indiscrete~(1)}

\item\textbf{nothing-to-hide~(10)}

\item\textbf{all-interests-correct~(8)}

\item\textbf{some-benefit~(8)}

\item\textbf{learned-something-new~(8)}

\emph{amount-of-information~(6), how-long-data-stored~(1)}

\item\textbf{interests-accurate~(8)}

\item\textbf{based-on-activities~(7)}

\emph{search~(5), maps~(2), work-related~(1)}

\item\textbf{only-few-interests~(6)}

\item\textbf{surprise~(5)}

\emph{amount-of-information~(4), how-old-data-was~(1)}

\item\textbf{does-not-trust-google~(5)}

\item\textbf{half-interests-correct~(4)}

\item\textbf{does-not-reflect-personality~(3)}

\item\textbf{reflects-personality~(3)}

\item\textbf{interests-too-generic~(2)}

\item\textbf{unsure~(2)}

\item\textbf{changed-perspective~(2)}

\item\textbf{feels-uncomfortable~(2)}

\item\textbf{missing-interests~(2)}

\item\textbf{physical-security-concerns~(2)}

\emph{location-data~(1), targeting-groups~(1)}

\item\textbf{surprised~(2)}

\item\textbf{google-know-me-better-than-I~(1)}

\item\textbf{repeated-search-metric~(1)}

\item\textbf{google-morally-ambiguous~(1)}

\item\textbf{reflects-history~(1)}

\emph{search~(1)}

\item\textbf{google-should-do-better~(1)}

\item\textbf{not-ashamed-of-activities~(1)}

\item\textbf{right-to-privacy~(1)}

\item\textbf{know-me-better-as-I~(1)}

\item\textbf{use-tools~(1)}

\item\textbf{not-surprised~(1)}

\item\textbf{google-products-necessary~(1)}

\item\textbf{companies-should-pay-for-data~(1)}

\item\textbf{predatory-advertising~(1)}

\item\textbf{benefits-not-worth-risk~(1)}

\item\textbf{left-google-services~(1)}

\item\textbf{order-incorrect~(1)}

\item\textbf{wants-to-learn-more~(1)}

\item\textbf{interests-derived-from-search~(1)}

\end{itemize}

 \end{small}

\end{document}